\documentclass[
	preprint,
	floats,floatfix,superscriptaddress,
	preprintnumbers,showpac,nofootinbib,dvipdfmx,
]{revtex4-1}
\usepackage{
	amsfonts,amsmath,epsfig,amssymb,latexsym,
	eucal,array,subfigure,bm,mathrsfs,
	ulem,epsfig,physics,color,graphicx,
	mathtools,
	}

\newcommand{\pd}{\partial}

\begin{document}

\title{Determining parameters of Kerr-Newman black holes \\ by shadow observation from finite distance and spatial infinity}
\author{Kenta Hioki}
\email{kenta.hioki@gmail.com}
\address{Sumitomo Mitsui Financial Group, Inc., 1-2, Marunouchi 1-chome, Chiyoda-ku, Tokyo 100-0005, Japan\footnote[1]{The statements expressed in this paper are those of the authors and do not represent the views of Sumitomo Mitsui Financial Group, Inc. or its staff.}}
\author{Umpei Miyamoto}
\email{umpei@akita-pu.ac.jp}
\address{Research and Education Center for Comprehensive Science, Akita Prefectural University, Akita 015-0055, Japan}
\begin{abstract}
We present a method for determining the physical parameters of a Kerr-Newman black hole through shadow observation. In a system comprising a Kerr-Newman black hole, an observer, and a light source, the relevant parameters are mass $M$, specific angular momentum $a$, electric charge $Q$, inclination angle $i$, and distance $r_o$. We consider the cases where the observer is at either a finite distance or spatial infinity. Using our method, the dimensionless parameters $(a/M, Q/M, i)$ can be determined by observing the shadow contour of the Kerr-Newman black hole from spatial infinity. We analytically prove that the shadow contour of the Kerr-Newman black hole observed from spatial infinity is unique, where uniqueness is defined as the absence of two congruent shadow contours for distinct sets of dimensionless parameter values. This method is versatile and can be applied to a range of black hole solutions with charge. Additionally, we show analytically that the shadow contour of a Kerr-Newman black hole observed from a finite distance $r_o$ is degenerate (not unique), meaning that the parameters of a Kerr-Newman black hole at finite distance cannot be determined from shadow observations.
\end{abstract}
\maketitle
\tableofcontents

\section{Introduction}
\label{sec:intro}
A black hole is a fascinating celestial object that serves as the ultimate testing ground for studying the physics of strong gravitational fields. Observing a black hole shadow has long been a significant goal in physics. This shadow is formed when light emitted by matter surrounding the black hole is bent by the black hole's intense gravitational pull.

Recently, Earth-sized very long baseline interferometers achieved a milestone by capturing images of two black hole candidates. The first image captured ${\rm M87}^\ast$~\cite{EventHorizonTelescope:2019dse, EventHorizonTelescope:2019uob, EventHorizonTelescope:2019jan, EventHorizonTelescope:2019ths, EventHorizonTelescope:2019ggy, EventHorizonTelescope:2021bee, EventHorizonTelescope:2021srq}, and the second focused on ${\rm Sgr~A}^\ast$~\cite{EventHorizonTelescope:2022xnr, EventHorizonTelescope:2022vjs, EventHorizonTelescope:2022wok, EventHorizonTelescope:2022exc, EventHorizonTelescope:2022urf, EventHorizonTelescope:2022xqj}. Both images displayed a prominent ring, marking a major advancement in black hole imaging techniques. However, the darker central region, known as the shadow, has yet to be clearly identified. Further enhancements in observation equipment are needed to resolve these shadows with greater precision~\cite{EventHorizonTelescope:2019ths}.

Some researchers suggest that the study of black hole imaging began with the derivation of the shadow contour, which is now known as the apparent shape~\cite{deVries:1999tiy}. The apparent shape of a Schwarzschild black hole was first derived in Ref.~\cite{darwin1959gravity}, followed by that of a Kerr black hole in Ref.~\cite{Bardeen:1973xx}. These foundational studies established the basis of shadow theory. Although these calculations focused on the apparent shapes of simple bare black holes without considering surrounding accretion disks, they highlighted the essential role of the photon sphere. This feature remains significant even in the shadow of a black hole with an accretion disk.

Many researchers have explored black hole shadows and methods for extracting physical information, such as a specific angular momentum of a black hole, through shadow observations~\cite{Hioki:2008zw, Bambi:2010hf, Amarilla:2010zq, Amarilla:2013sj, Wei:2013kza, Papnoi:2014aaa, Wei:2015dua, Singh:2017vfr, Stuchlik:2019uvf, Tsukamoto:2024gkz}. In this paper, however, we emphasize that there is still potential to improve the accuracy of measuring black hole parameters via shadow analysis. In particular, further study is needed to determine whether information like a specific angular momentum and electric charge of a black hole can be obtained solely from shadow observations.

One way to investigate this possibility is examining whether the map from a parameter space to an image library is injective~\cite{Hioki:2009na, Hioki:2022mdg}. It has been shown that the map from the parameter space to the apparent-shape library for a bare Kerr black hole is indeed injective~\cite{Hioki:2009na, Hioki:2023ozd}. Here, the apparent-shape library refers to the set of all possible apparent shapes that can be produced within the framework of the given gravitational theory and model. The result suggests that the dimensionless specific angular momentum and inclination angle of a Kerr black hole can be uniquely determined by observing its apparent shape. 

Determining (1) whether an object is indeed a black hole, (2) identifying the specific black hole solution, and (3) accurately measuring its physical parameters based solely on shadow images is a challenging problem that requires extensive research. It is essential to generate a wide range of models for relativistic objects, thereby building an apparent-shape library that incorporates diverse parameters. Moreover, investigating the injectivity of the map is critical. If the map is not injective, it indicates the presence of images that correspond to different models and parameter configurations, making it impossible to uniquely determine the model based on shadow observations alone. These research efforts are ongoing, and further studies are necessary~\cite{Paganini:2017qfo, Mars:2017jkk, Lima:2021las}.

To address this problem, we assume that the black hole solution of the observed object has been identified, and it must be demonstrated that its parameters can be determined from shadow observations. When the target object is a black hole with three physical parameters (mass, angular momentum, and charge), no method has yet been proposed to uniquely determine these parameters from shadow observations alone, rather than merely constraining them.

For this purpose, we assume that the observed object is a Kerr-Newman black hole. The system comprising the Kerr-Newman black hole and the observer includes the physical quantities of mass $M$, specific angular momentum $a$, electric charge $Q$, inclination angle $i$, and distance $r_o$. We consider both cases in which the observer is at a finite distance and at spatial infinity.

For the first time in this paper, we show with a concrete method that the dimensionless parameters $(a/M, Q/M, i)$ can be uniquely determined by observing the shadow when the observer is at spatial infinity. The key point is that the aforementioned map from the parameter space to the image library is injective. This method can be applied to various black hole solutions with charge and is highly versatile.

We also show analytically that there is uniqueness in the contour of the black hole shadow when the observer is at spatial infinity. Uniqueness is defined as the absence of two congruent apparent shapes for two distinct dimensionless parameter values. The uniqueness of the apparent shape is the basis of our method for determining dimensionless parameters from shadow observations. 

Surprisingly, we analytically prove that uniqueness does not hold for the shadow contour of the Kerr-Newman black hole when the distance $r_{o}$ of the black hole is finite. This is a different feature from the case of the apparent shape of the Kerr black hole at finite distance~\cite{Hioki:2023ozd}. This will be a problem to be solved in the future when there is an opportunity to observe black holes from a finite distance. It would be interesting to consider what would happen to this point if a more realistic model were considered, such as a black hole with an accretion disk.

The structure of this paper is as follows. In Sec.~\ref{sec:setup}, we begin the analysis by deriving the null geodesic equations around the Kerr-Newman black hole and describing how to construct an apparent shape from these geodesics. In Sec.~\ref{sec:shadow}, we present examples of the apparent shapes of the Kerr-Newman black hole at finite distances and then analytically demonstrate the non-uniqueness of these shapes. In Sec.~\ref{sec:pc}, we analytically prove the uniqueness of the apparent shape in Bardeen coordinates. We also propose a method to systematically construct observables that characterize the apparent shape of Kerr-Newman black holes, showing that the dimensionless parameters of the system can be determined from shadow observations. In the final section, we summarize our analysis and discuss future prospects. We use geometrized units, where $c=G=1$.

\section{Setup}
\label{sec:setup}
\subsection{Null geodesics}
The Kerr-Newman spacetime is a stationary, axisymmetric, and asymptotically flat solution of the Einstein-Maxwell theory~\cite{Newman:1965my}. The Kerr-Newman metric in Boyer-Lindquist coordinates has the form
\begin{eqnarray}
	g_{\mu \nu}{\rm d}x^\mu {\rm d}x^\nu 
	=
	&-&\left(
		1-\frac{2Mr - Q^2}{\varSigma}
	\right) {\rm d}t^2
	+
	\frac{\varSigma}{\varDelta} {\rm d}r^2
	+
	\varSigma {\rm d}\theta ^2 \nonumber \\
	&-&
	\frac{ \left( 2Mr - Q^2 \right) 2a \sin^2\theta }{ \varSigma } {\rm d}t {\rm d}\phi
	+
	\frac{A \sin ^2 \theta}{\varSigma} {\rm d}\phi ^2
	,
	\label{eq:metric}
\end{eqnarray}
where $x^\mu = \left( t, r, \theta , \phi \right)$ $\left( \mu , \nu = 0, 1, 2, 3 \right)$,
\begin{eqnarray}
	\varSigma (r,\theta)
	&:=&
	r^2 + a^2 \cos^2 \theta ,
	\nonumber \\
	\varDelta (r)
	&:=&
	r^2 - 2Mr + a^2 + Q^2 ,
	\nonumber \\
	A (r,\theta)
	&:=&
	\left( r^2 + a^2 \right) ^2 - a^2 \varDelta \sin^2\theta.
\end{eqnarray}
The parameters $M$, $a$, and $Q$ are the mass, specific angular momentum, and electric charge of the spacetime, respectively. If $0 \leq a^2 + Q^2 \leq M^2$, then an event horizon exists in the spacetime, and the metric describes a black hole.
The radii of the outer and inner horizons are denoted by $r_{+}$ and $r_{-}$, respectively.
The Kerr-Newman metric contains Kerr $\left( a \neq Q = 0 \right)$, Reissner-Nordstr${\rm \ddot{o}}$m $\left( a = 0 \neq Q \right)$, and Schwarzschild $\left( a = Q = 0 \right)$ metrics as special cases.

We shall solve the null geodesic equation to find the trajectory $x^\mu \left( \lambda \right)$ of a massless test particle. We have four independent constants of motion in involution for geodesics in the Kerr-Newman black hole, making the geodesic equation completely integrable~~\cite{Frolov:1998wf}.

The Lagrangian of the massless test particle in the spacetime is expressed as:
\begin{eqnarray}
	\mathcal{L} &=& \frac{1}{2}g_{\mu \nu}\dot{x}^\mu \dot{x}^\nu \, , \\
	\dot{x}^\mu &\coloneqq& \frac{dx^\mu}{d\lambda} \, .
	\label{eq:lag}
\end{eqnarray}
We derive the energy $E$ and the axial component of the angular momentum $L$ of the test particle:
\begin{eqnarray}
	E \coloneqq  - \frac{\partial \mathcal{L}}{\partial \dot{t}} = \left( 1-\frac{2Mr - Q^2}{\varSigma} \right) \dot{t} +\frac{ \left( 2Mr - Q^2 \right) 2a \sin^2\theta }{ \varSigma } \dot{\phi}
\end{eqnarray}
and
\begin{eqnarray}
	L \coloneqq  \frac{\partial \mathcal{L}}{\partial \dot{\phi}}  = -\frac{ \left( 2Mr - Q^2 \right) 2a \sin^2\theta }{ \varSigma } \dot{t} + \frac{A \sin ^2 \theta}{\varSigma} \dot{\phi} \, .
\end{eqnarray}
These quantities $\mathcal{L}$, $E$, and $L$, along with $\mathcal{Q}$ (known as the Carter constant~\cite{Chandrasekhar:1985kt, Frolov:1998wf}), are conserved, highlighting their roles as constants of motion.

We denote the four-momentum of a massless test particle by $k^\mu$, where
\begin{eqnarray}
	k^\mu = \frac{{\rm d} x^\mu}{{\rm d}\tilde{\lambda}} \, .
\end{eqnarray}
Here, $\tilde{\lambda}$, defined as $\tilde{\lambda} \coloneqq \lambda E$, serves as the affine parameter. We introduce two conserved quantities for null geodesics $(\mathcal{L} = 0)$: $\ell \coloneqq L/E$ and $\mathbb{Q} \coloneqq \mathcal{Q}/E^2$. Consequently, this leads us to obtain the following set of first-order differential equations~\cite{Frolov:1998wf}:
\begin{eqnarray}
	&&
 	\varSigma k^t
	=
	\frac{A - a \ell \left( 2Mr -Q^2 \right) }{\varDelta} \, ,
	\label{eq:velocity2}	
	\\
	&&
	\varSigma k^r
	=
	\pm \sqrt{R} \, ,
	\label{eq:velocity0}
	\\
	&&
	\varSigma k^\theta
	=
	\pm \sqrt{\varTheta} \, ,
	\label{eq:velocity1}
	\\
	&&
	\varSigma k^\phi
	=
	\frac{ a \left( 2Mr - Q^2 \right) + \ell \csc ^2\theta \left( \varSigma - 2Mr + Q^2 \right) }{\varDelta} \, ,
	\label{eq:velocity3}
\end{eqnarray}
where
\begin{eqnarray}
	K  &\coloneqq& \mathbb{Q} +\left( a-\ell \right) ^2 \, ,
	\label{eq:cqk}
	\\
	R(r)
	&\coloneqq&
        \left( r^2+a^2-a\ell \right) ^2- K \varDelta \, , \\
	\varTheta(\theta)
	&\coloneqq&
	K -(a \sin \theta -\ell \csc \theta )^2.
\end{eqnarray}

\subsection{Light source and observer}
If the initial conditions are set, the trajectory of the massless test particle can be found according to the geodesic equation.

We consider the setting of the light source. For astronomical black holes such as ${\rm M87}^\ast$ to be the object of observation, let us assume that the black hole has neither an accretion flow nor an accretion disc surrounding it~\cite{EventHorizonTelescope:2021dqv, Vagnozzi:2022moj}. This means that we assume that the Kerr-Newman black hole is bare. In the next paragraphs, we will explain our assumptions about the observer in more detail, where we assume that the distance between the black hole and the observer takes a finite value $r_o$. Then, we assume that a sphere of radius $r=r_e = {\rm const.}$, where $r_o < r_e$, is the light source and that each point on it uniformly emits a null ray~\cite{Grenzebach:2014fha}.

Next, let us clarify the setting of the observer receiving the null ray. Possible candidates include zero-angular-momentum observers \cite{Bardeen:1973xx} and Carter's observers \cite{Grenzebach:2014fha}, each of which exhibits distinct azimuthal motions \cite{Chang:2020lmg}. Our aim is to demonstrate the existence of an observer capable of determining the black hole parameters. Selecting either observer type is sufficient for this purpose. In this paper, we choose Carter's observer.

We introduce a tetrad of basis vectors,
\begin{eqnarray}
	e_{(t)} &\coloneqq& \frac{\left( r^2 + a^2 \right) \pd_t + a \pd_\phi }{\sqrt{\varSigma \varDelta}} \, , \\
	e_{(r)} &\coloneqq& - \sqrt{\frac{\varDelta}{\varSigma}} \pd_r \, ,  \\
	e_{(\theta)} &\coloneqq& \frac{1}{\sqrt{\varSigma}} \pd_\theta \, , \\
	e_{(\phi)} &\coloneqq& -\frac{ a \sin \theta \pd_t + \csc \theta \pd_\phi }{\sqrt{\varSigma}}  \, .
\label{eq:tetrad}
\end{eqnarray}
The timelike vector $e_{(t)} | _{(r, \theta ) = (r_o, i)}$ can be interpreted as the four-velocity of the observer and the vector $e_{(r)} | _{(r, \theta ) = (r_o, i)}$ gives the spatial direction towards the black hole [see Fig.~\ref{fig010}(a)]. Since Eqs.~(\ref{eq:velocity1}) and (\ref{eq:velocity3}) are defined on $\theta \in (0, \pi)$, we assume $i \in (0 , \pi/2 ]$.

We assume that the observer is at finite distance $r_o$ from the black hole. 
While the observer could be located anywhere within the domain of outer communication, we limit this range to $r_o \in [5M, \infty ]$. This restriction is adequate for demonstrating that the black hole's parameters can be determined.

\subsection{Celestial coordinates and screen coordinates}
\begin{figure}[tb]
		\begin{tabular}{ cc }
			\includegraphics[height=5.5cm]{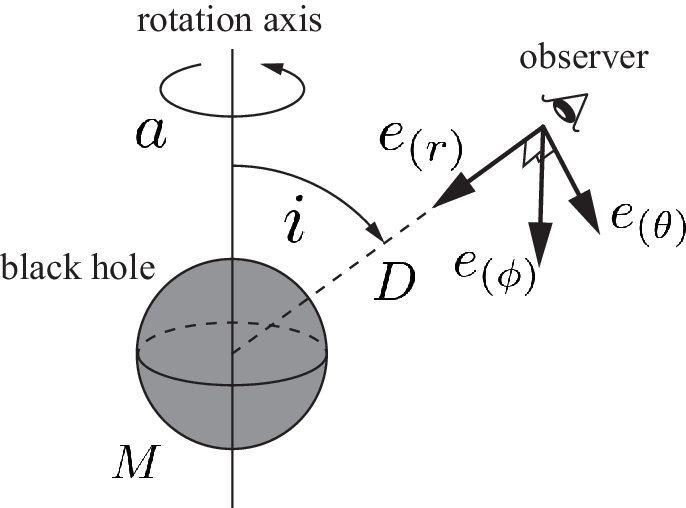} &
			\includegraphics[height=5.5cm]{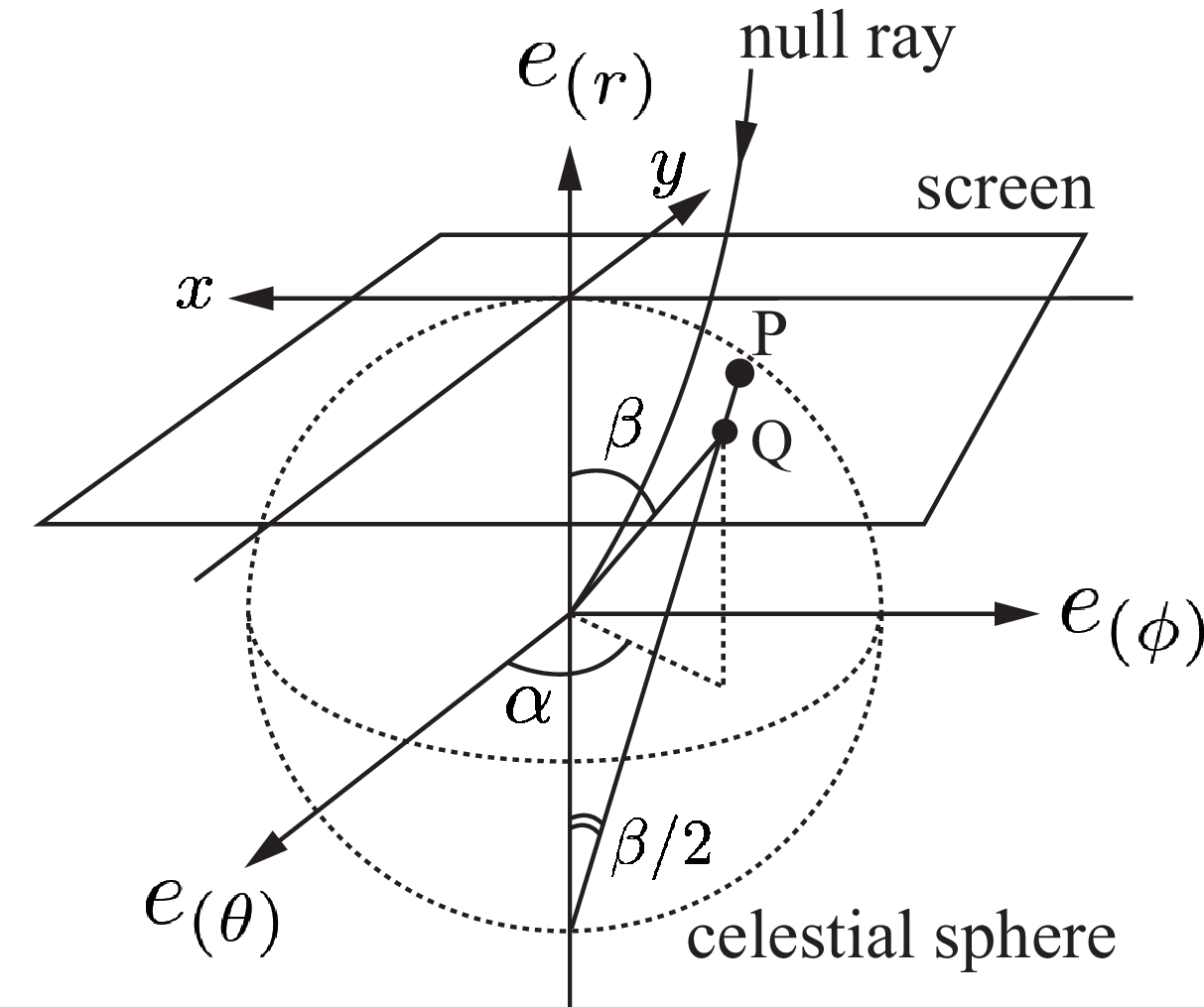}
   \\
			(a) & (b)
		\end{tabular}
	\caption{\footnotesize{
(a) A schematic picture showing a unit sphere in which the center coincides with that of the Kerr-Newman black hole. The observer is located at $(r, \theta) = (r_o, i)$. (b) A schematic picture showing a unit celestial sphere for the observer. The direction of $e_{(r)}$ is a direction to the Kerr-Newman black hole from the observer. $(\alpha , \beta)$ is a celestial coordinate system, which is used to specify the incident angle of the null ray into the observer. Q is the point where the tangent of null ray at the center of the celestial sphere crosses the unit sphere. P is the two-dimensional image made by the null ray.}}
	\label{fig010}
\end{figure}
In the remaining subsections of Sec.~\ref{sec:setup}, we formulate how the apparent shape, which is the contour of the shadow, is formed by null rays. Specifically, we present the map from the parameters to the apparent shape.

The tangent vector of the null geodesic $x^\mu(\tilde{\lambda})$ can be written, using four-momentum $k^\mu$, as follows:
\begin{eqnarray}
	\frac{{\rm d}}{{\rm d} \tilde{\lambda}} = k^\mu \partial _\mu \, .
\label{eq:tangentv1}
\end{eqnarray}
As can be seen from Fig.~\ref{fig010}(b), this tangent vector at the observer can be expressed using the two incident angles $(\alpha , \beta)$ of the null geodesic to the observer as follows:
\begin{eqnarray}
	\left. \frac{{\rm d}}{{\rm d} \tilde{\lambda}} \right|_{(r,\theta)=(r_o,i)} = \epsilon \left( - e_{(t)} + \cos \beta e_{(r)} + \sin \beta \cos \alpha e_{(\theta)} + \sin \beta \sin \alpha e_{(\phi)} \right) ,
	\label{eq:tangentv2}
\end{eqnarray}
where $\epsilon$ is a scalar factor. We refer to these two incident angles $(\alpha , \beta)$ as the celestial coordinates of the observer. The scale factor $\epsilon$ can be derived from the orthogonality of the basis vectors:
\begin{eqnarray}
	\epsilon = - \left. \frac{ r^2 + a^2 - a\ell}{\sqrt{\varSigma \varDelta}} \right| _{(r, \theta) = (r_o, i)} \, .
	\label{def_epsilon}
\end{eqnarray}
Using Eqs.~(\ref{eq:tangentv1}), (\ref{eq:tangentv2}), and (\ref{def_epsilon}), we write down the equations relating the celestial coordinates, constants of motion and parameters~\cite{Grenzebach:2014fha}.
\begin{eqnarray}
	&&
	\sin \alpha
	=
	\left. \frac{\sin \theta}{\sqrt{\varDelta}\sin \beta} \left( \frac{\Delta \varSigma k^\phi}{r^2 + a^2  - a \ell} -a \right) \right| _{(r, \theta) = (r_o, i)} \, ,
	\label{eq:alpha}
	\\
	&&
	\cos \beta
	=
	\left. \frac{\varSigma k^r }{r^2 + a^2 - a \ell} \right| _{(r, \theta) = (r_o, i)} \, .
	\label{eq:beta}
\end{eqnarray}

Let us define the screen coordinates $(x, y)$ in which the apparent shape of the black hole is depicted, as shown in Fig.~\ref{fig010}(b)~\cite{Grenzebach:2014fha}. The celestial coordinates are transformed to the screen coordinates as follows:
\begin{eqnarray}
	x
	&=&
	-2 \sin \alpha \tan \frac{\beta}{2}  \, ,
	\label{eq:x}\\
	y
	&=&
	-2 \cos \alpha \tan \frac{\beta}{2} \, .
	\label{eq:y}
\end{eqnarray}
A circle with center at the origin and radius 2 in the screen coordinates $(x, y)$ corresponds to the celestial equator.

\subsection{Classes of null geodesics}
We consider two important classes of null geodesics forming the apparent shape. The first is the class of spherical photon orbits, and the second is the class of principal null geodesics.

Whether a null geodesic emitted from a light source has a turning point after heading toward a black hole and reaches the observer depends on the conserved quantities of the null geodesic. The critical value of the conserved quantities that determines this is the value of the conserved quantities of the null geodesic that winds around the spherical photon orbit an infinite number of times. We can say that the value of the conserved quantities of the spherical photon orbits is critical. This is because spherical photon orbits and the null geodesics that wind around those orbits are null geodesics with different initial conditions, but with the same conserved quantity values.

A null geodesic with constant radial motion is called a {\it spherical photon orbit}. The radius $r_s$ of a spherical photon orbit is obtained by~\cite{deVries:1999tiy, Takahashi:2005hy}
\begin{eqnarray}
	R(r_s)
	 = 
	0  \, ,
\;\;\;
	\frac{{\rm d}R(r_s)}{{\rm d}r_s}
	 = 
	0 \, .
	\label{eq:spo}
\end{eqnarray}
For a spherical photon orbit of radius $r_s$ to exist, there is an additional condition that $\theta$ must satisfy the following inequality
\begin{eqnarray}
	\varTheta (\theta)
	& \geq &
	0  \, .
	\label{eq:contheta}
\end{eqnarray}
The conserved quantities $\ell$ and $\mathbb{Q}$ satisfying Eq.~\eqref{eq:spo} are denoted by $\ell _s$ and $\mathbb{Q}_s$, where
\begin{eqnarray}
	\ell _s
	&=& 
	\frac{r_s^2+a^2}{a} - \frac{2r_s\varDelta (r_s)}{a(r_s-M)} \, ,
	\label{eq:spoell}
	\\
	\mathbb{Q} _s
	&=&
	- \frac{r_s^2 \left[r_s^2 \left(r_s-3M \right) ^2-4M r_s \left(a^2 +Q^2 \right)+4Q^2 \Delta \right]}{a^2(r_s-M)^2} \, ,
	\label{eq:spoq}
\end{eqnarray}
respectively~\cite{deVries:1999tiy, Takahashi:2005hy}. These quantities $\ell _s$ and $\mathbb{Q}_s$ are conserved quantities for spherical photon orbits of radius $r_s$.
Considering the position of the observer, it can be seen from Eqs.~(\ref{eq:contheta}), (\ref{eq:spoell}), and (\ref{eq:spoq}) that the radius $r_s$ must satisfy $\varTheta (i) | _{(\ell , \mathbb{Q}) = (\ell _s , \mathbb{Q}_s)} \geq 0$.

Substituting Eqs.~(\ref{eq:spoell}) and (\ref{eq:spoq}) into Eq.~(\ref{eq:cqk}), we find that the conserved quantity $K$ of the spherical photon orbits is
\begin{eqnarray}
	K=K_s:= \frac{4r_s^2\varDelta(r_s) }{(r_s - M)^2}  \, .
	\label{eq:cartercnst}
\end{eqnarray}
Since the conserved quantity $K$ is generally non-negative \cite{Chandrasekhar:1985kt}, the radius $r_s$ of a spherical photon orbit must lie within the range
\begin{eqnarray}
	r_s \in ( -\infty , r_{-}] \cup [r_{+}, \infty ) \, .
	\label{eq:spoexist}
\end{eqnarray}
This is a necessary condition for the existence of spherical photon orbits of radius $r_s$.

The condition under which the spherical photon orbit is unstable in the radial direction is
\begin{eqnarray}
	\frac{{\rm d}^2R(r_s)}{{\rm d}r_s^2} > 0 \, .
	\label{eq:sta}
\end{eqnarray}
This inequality is equivalent to 
\begin{eqnarray}
	r_s \in ( -\infty , 0 ) \cup \left( r_u, \infty \right) \, ,
	\label{eq:spounsta}
\end{eqnarray}
where $r_u \coloneqq M-\left[ M \left( M^2-a^2-Q^2 \right) \right]^{1/3}$. Here, $r_u$ is the non-negative real root of the function ${\rm d}^2R(r_s)/{\rm d}r_s^2$. The function ${\rm d}^2R(r_s)/{\rm d}r_s^2$ has at most two real roots: $r_u$ and $0$.

As can be seen from the conditions~(\ref{eq:spoexist}) and (\ref{eq:spounsta}), if an unstable spherical photon orbit exists, its radius $r_s$ satisfies the following:
\begin{eqnarray}
	r_{s} \in ( -\infty , 0 ) \cup [ r_{+} , \infty  )  \, .
	\label{eq:rsta}
\end{eqnarray}
Thus, all spherical photon orbits outside the event horizon that are important for the apparent shape are unstable in the radial direction.
For the sake of later discussion, we define $I$ as the set of radii of the existing unstable spherical photon orbits.

Another important class is that of {\it principal null geodesics}~\cite{Chandrasekhar:1985kt, Frolov:1998wf}. The principal null geodesics reach the observer straight $\left( \theta = {\rm const.} \right)$ from the black hole, without any turning points in the radial direction.

The principal null geodesic is the null geodesic for which the conserved quantity $K$ is $K=0$. When we assume that the principal null geodesic reaches the observer, Eq.~(\ref{eq:velocity1}) implies that $\theta = i$. These give us two conserved quantities $\ell _p$ and $\mathbb{Q} _p$ of the principal null geodesic:
\begin{eqnarray}
	\ell _p &=& a \sin ^2 i \, , \\
	\mathbb{Q} _p &=& - a^2 \cos ^4 i   \, .
\end{eqnarray}
The origin of the screen coordinates defined in Fig.~\ref{fig010} is related to the conserved quantity $\ell _p$ of the principal null geodesics. This point will be discussed in more detail in Sec.~\ref{sec:pc}.

\subsection{Shadow of the black hole}
We describe how the apparent shape, the shadow contour of a black hole, can be drawn. To do so, we describe the role of the unstable spherical photon orbits and the principal null geodesics in the casting of the shadow.

To consider unstable spherical photon orbits, let us trace the geodesics reaching the observer back to their starting points.
Null geodesics that do not start from a light source are null geodesics forming the shadow of a black hole. Such geodesics can be classified into two categories. One is a null geodesic whose starting point is a black hole. The other is a null geodesic that is winding around a spherical photon orbit an infinite number of times.

We shall draw a curve $c$ in screen coordinates corresponding to a geodesic winding around a spherical photon orbit an infinite number of times. The functions $x(\ell _s, \mathbb{Q}_s)$ and $y(\ell _s, \mathbb{Q}_s)$ can be obtained by substituting Eqs.~(\ref{eq:velocity0}), (\ref{eq:velocity3}), (\ref{eq:alpha}), and (\ref{eq:beta}) into Eqs.~(\ref{eq:x}) and (\ref{eq:y}). Then, using Eqs.~(\ref{eq:spoell}) and (\ref{eq:spoq}), the functions become $x(r_s)$ and $y(r_s)$. We can draw a closed curve $c$ in the screen coordinates for fixed values of $(M, r_o, a, Q, i)$ by varying $r_s$ over the entire range of $I$.

The closed curve $c$ is the boundary between the dim and bright areas on the screen coordinates~\cite{deVries:1999tiy, Hioki:2008zw, Hioki:2009na}. Intuitively, the inside of the closed curve is considered to be the dark area, but let us briefly explain this using the principal null geodesic. Since the principal null geodesic reaches the observer from the black hole, it is the point that constitutes the dim part on the screen coordinates. Since this point is located inside the closed curve, the interior of the closed curve is the dim part, the shadow~\cite{Hioki:2009na}. Of course, the closed curve itself also constitutes part of the shadow. Thus, the closed curve $c$ is the apparent shape of the black hole.

\section{Properties of shadows}
\label{sec:shadow}
\subsection{Examples of apparent shapes}
\label{subsec:examples}
\begin{figure}[tb]
		\begin{tabular}{ cccc }
			\includegraphics[height=3.8cm]{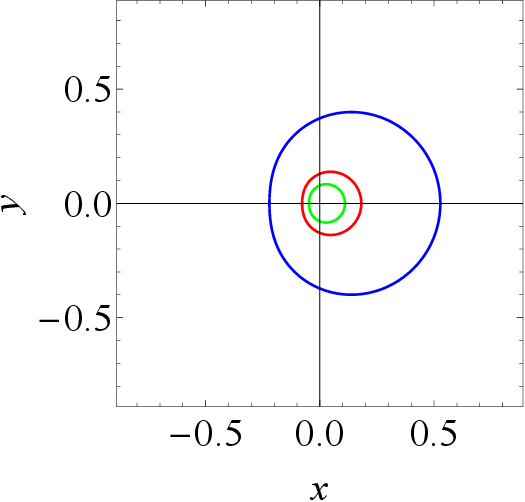} &
			\includegraphics[height=3.8cm]{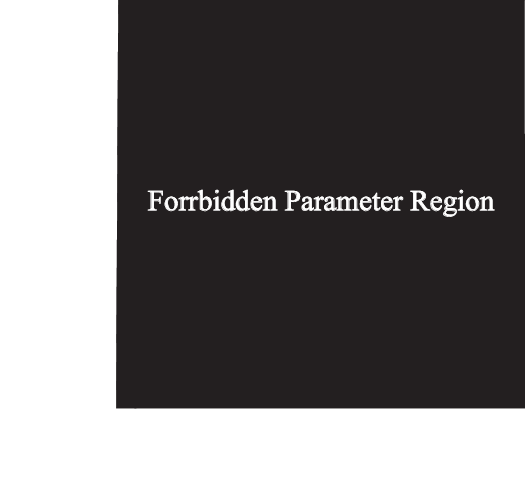} &
			\includegraphics[height=3.8cm]{figs/NS.eps} &
			\includegraphics[height=3.8cm]{figs/NS.eps}
			\\
			\scriptsize{(a) $a/M=0.4$, $Q/M=0.9165$} & & & \\
			\includegraphics[height=3.8cm]{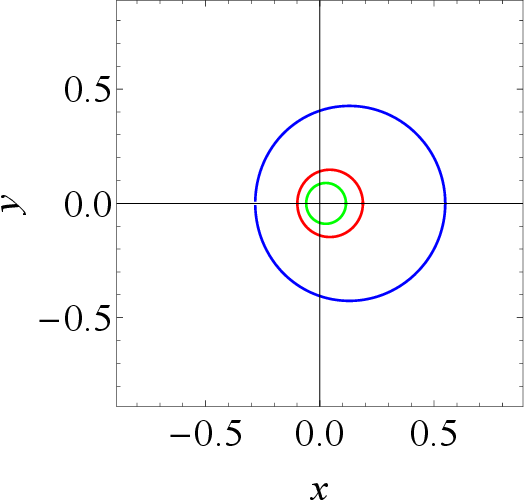} &
			\includegraphics[height=3.8cm]{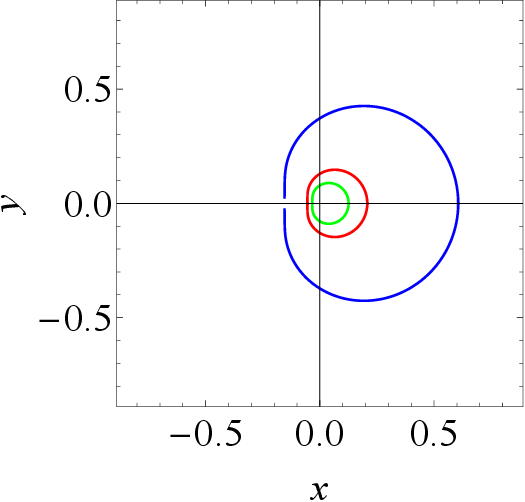} &
			\includegraphics[height=3.8cm]{figs/NS.eps} &
			\includegraphics[height=3.8cm]{figs/NS.eps}
			\\
			\scriptsize{(b) $a/M=0.4$, $Q/M=0.7999$} & \scriptsize{(c) $a/M=0.6$, $Q/M=0.7999$} & & \\
			\includegraphics[height=3.8cm]{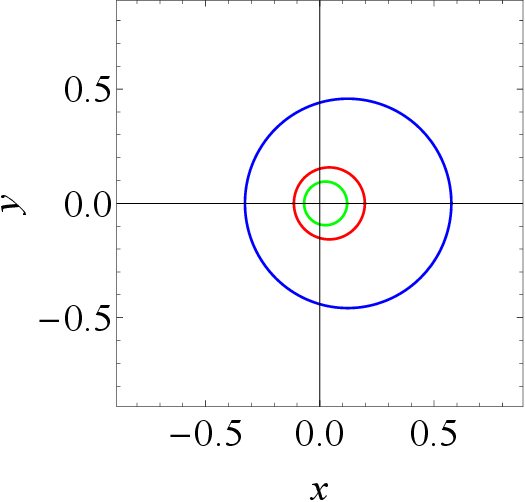} &
			\includegraphics[height=3.8cm]{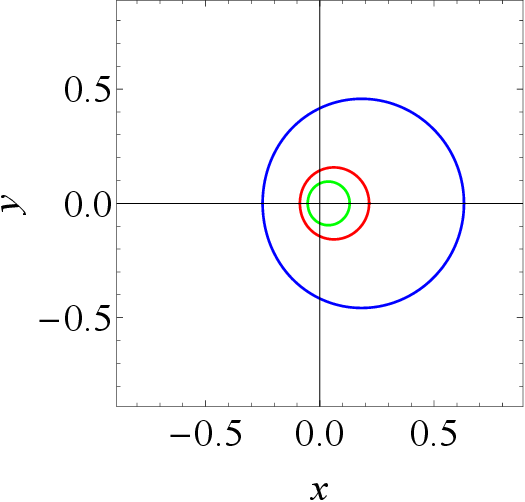} &
			\includegraphics[height=3.8cm]{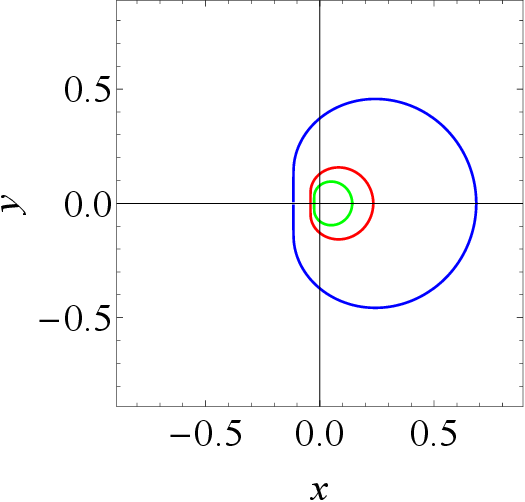} &
			\includegraphics[height=3.8cm]{figs/NS.eps}
			\\
			\scriptsize{(d) $a/M=0.4$, $Q/M=0.5999$} & \scriptsize{(e) $a/M=0.6$, $Q/M=0.5999$} & \scriptsize{(f) $a/M=0.8$, $Q/M=0.5999$} & \\
			\includegraphics[height=3.8cm]{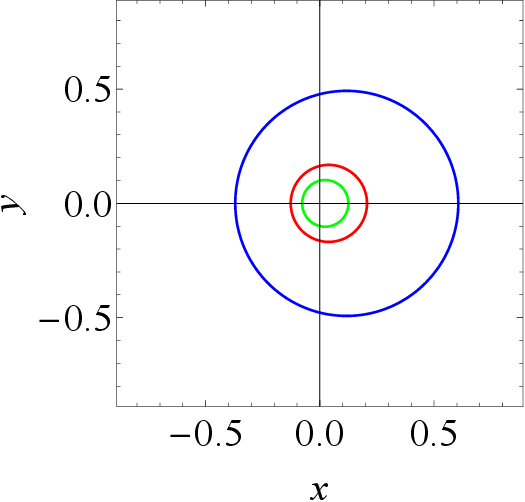} &
			\includegraphics[height=3.8cm]{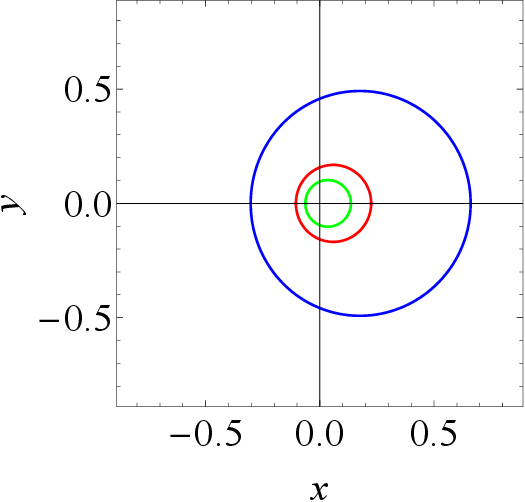} &
			\includegraphics[height=3.8cm]{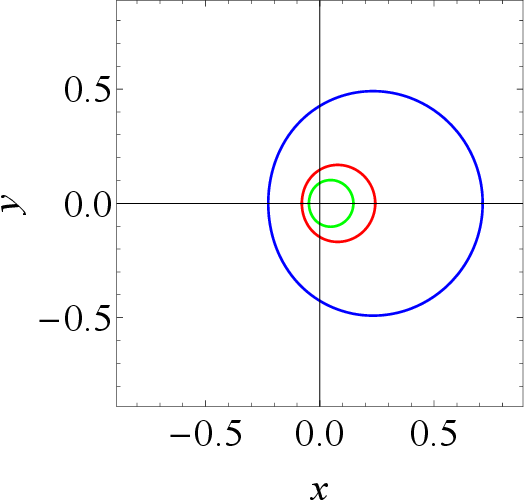} &
			\includegraphics[height=3.8cm]{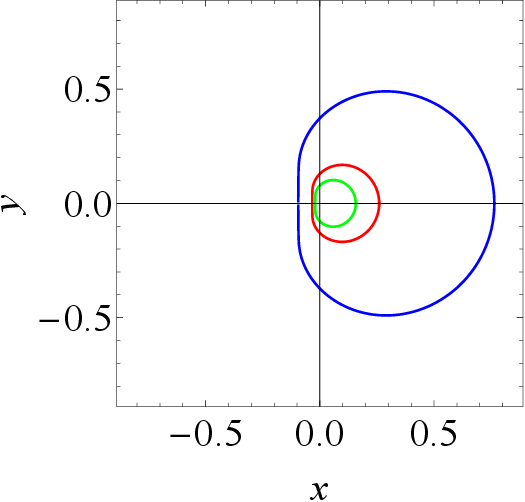} \\
			\scriptsize{(g) $a/M=0.4$, $Q/M=0.04$} & \scriptsize{(h) $a/M=0.6$, $Q/M=0.04$} & \scriptsize{(i) $a/M=0.8$, $Q/M=0.04$} & \scriptsize{(j) $a/M=0.999$, $Q/M=0.04$}
		\end{tabular}
	\caption{\footnotesize{Shadows of the Kerr-Newman black holes with the inclination angle $i =  90^\circ$. The blue, red, and green curves are the apparent shape of the black hole at dimensionless distances $r_o/M$ of 10, 30, and 50, respectively.}}
	\label{fig020}
\end{figure}
\begin{figure}[tb]
		\begin{tabular}{ cccc }
			\includegraphics[height=3.8cm]{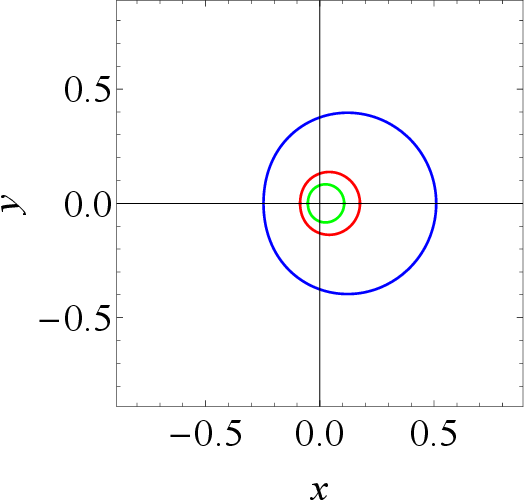} &
			\includegraphics[height=3.8cm]{figs/NS.eps} &
			\includegraphics[height=3.8cm]{figs/NS.eps} &
			\includegraphics[height=3.8cm]{figs/NS.eps}
			\\
			\scriptsize{(a) $a/M=0.4$, $Q/M=0.9165$} & & & \\
			\includegraphics[height=3.8cm]{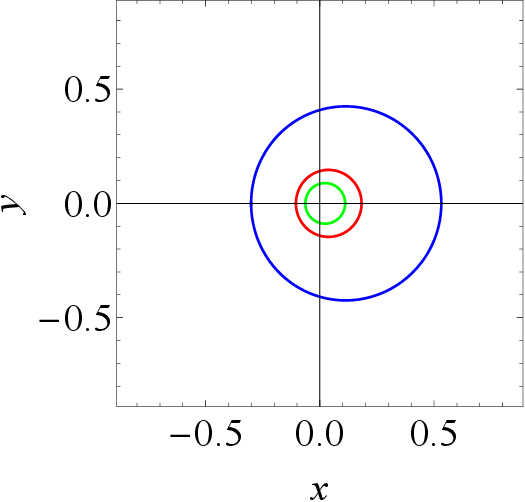} &
			\includegraphics[height=3.8cm]{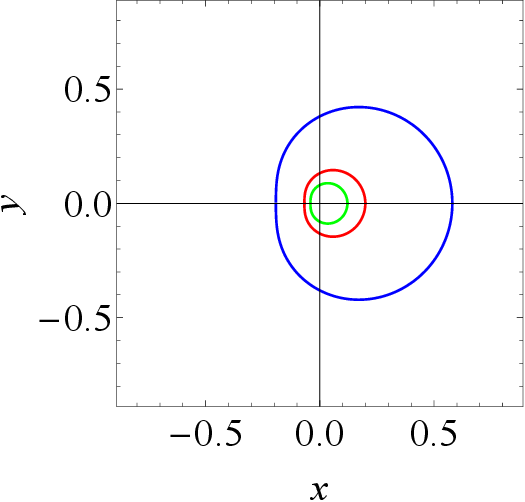} &
			\includegraphics[height=3.8cm]{figs/NS.eps} &
			\includegraphics[height=3.8cm]{figs/NS.eps}
			\\
			\scriptsize{(b) $a/M=0.4$, $Q/M=0.7999$} & \scriptsize{(c) $a/M=0.6$, $Q/M=0.7999$} & & \\
			\includegraphics[height=3.8cm]{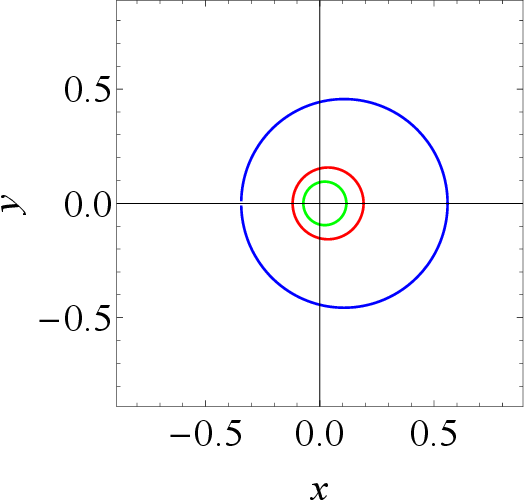} &
			\includegraphics[height=3.8cm]{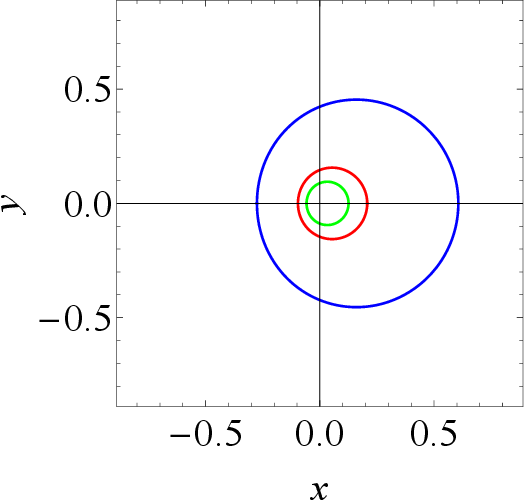} &
			\includegraphics[height=3.8cm]{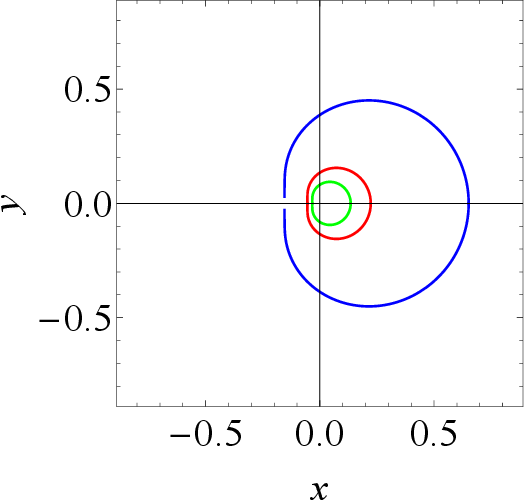} &
			\includegraphics[height=3.8cm]{figs/NS.eps}
			\\
			\scriptsize{(d) $a/M=0.4$, $Q/M=0.5999$} & \scriptsize{(e) $a/M=0.6$, $Q/M=0.5999$} & \scriptsize{(f) $a/M=0.8$, $Q/M=0.5999$} & \\
			\includegraphics[height=3.8cm]{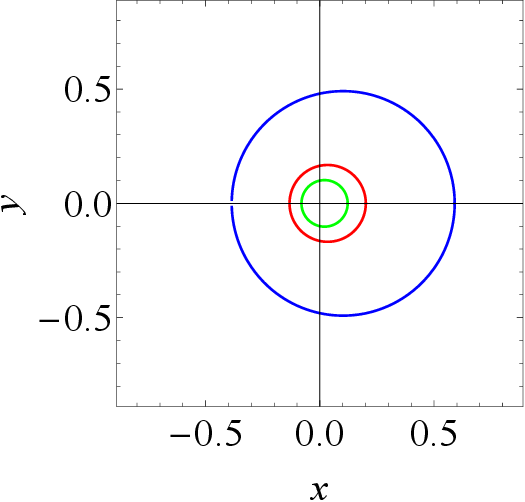} &
			\includegraphics[height=3.8cm]{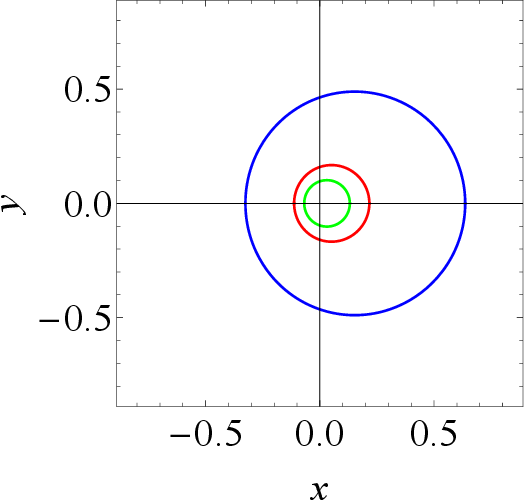} &
			\includegraphics[height=3.8cm]{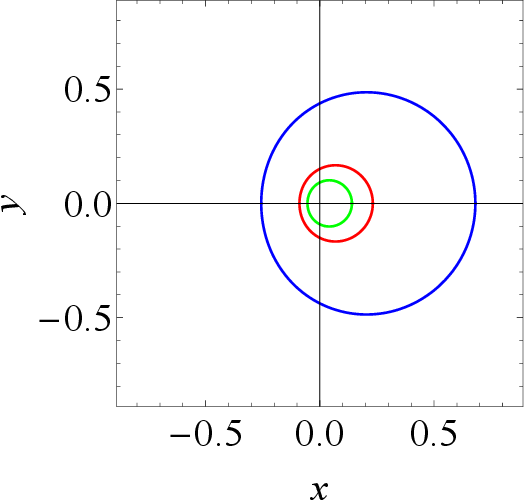} &
			\includegraphics[height=3.8cm]{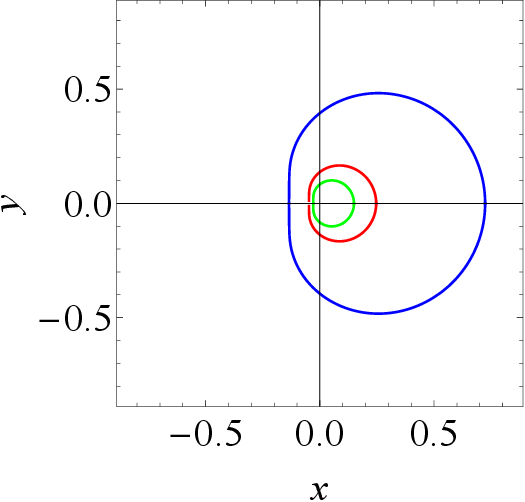} \\
			\scriptsize{(g) $a/M=0.4$, $Q/M=0.04$} & \scriptsize{(h) $a/M=0.6$, $Q/M=0.04$} & \scriptsize{(i) $a/M=0.8$, $Q/M=0.04$} & \scriptsize{(j) $a/M=0.999$, $Q/M=0.04$}
		\end{tabular}
	\caption{\footnotesize{Shadows of the Kerr-Newman black holes with the inclination angle $i =  60^\circ$. The blue, red, and green curves are the apparent shape of the black hole at dimensionless distances $r_o/M$ of 10, 30, and 50, respectively.}}
	\label{fig025}
\end{figure}
In Figs.~\ref{fig020} and \ref{fig025}, we present the apparent shapes of the Kerr-Newman black holes for four values of dimensionless spin parameter $a/M = 0.4$, 0.6, 0.8, and 0.999, four values of dimensionless electric charge $Q/M = 0.04$, 0.5999, 0.7999, and 0.9165, three values of dimensionless distance $r_o/M = 10$, 30, and 50, and two values of inclination angle $i = 90^\circ $ and $60^\circ $. It can be seen from Figs~\ref{fig020} and \ref{fig025}. that the deformation of the apparent shape depends mainly on $a/M$ and $Q/M$, $i$, and that the size depends mainly on $r_o/M$. We will see in Sec.~\ref{sec:pc} that the parameter $r_o/M$ also deform the apparent shape. Note that the apparent shapes with different $r_o/M$ but the same values of other parameters are {\it not similar}. Namely, we mean that one shape cannot be obtained from the other by uniform scaling, possibly with translation, rotation, or reflection.

\subsection{Degeneracy of apparent shape of the black hole at finite distance}
\label{sec:uniq1}
To find out the characteristics of the apparent shape, we need to analyze the corresponding curves on the celestial coordinates $(\alpha , \beta )$. As can be seen from Eqs. (\ref{eq:alpha}) and (\ref{eq:beta}), the functions $\sin \alpha$ and $\cos \beta$ are not explicitly dependent on $M$. We introduce dimensionless parameters:
\begin{eqnarray}
	r_{o\ast} \coloneqq \frac{r_o}{M}, \;\;\;
	a_\ast \coloneqq \frac{a}{M}, \;\;\;
	Q_\ast \coloneqq \frac{Q}{M}, \;\;\;
	r_{\ast} \coloneqq \frac{r_s}{M}.
	\label{eq:Dar}
\end{eqnarray}

We prove the degeneracy (non-uniqueness) of the apparent shape on the screen coordinates of the Kerr-Newman black hole. We refer to degeneracy and uniqueness as the existence and absence of two congruent apparent shapes at the same position on the screen coordinates for two values of parameters $\left( r_{o\ast} , a_\ast , Q_\ast , i \right)$, respectively~\cite{Mars:2017jkk, Lima:2021las, Hioki:2023ozd}.

To verify whether the apparent shape on the screen coordinates is uniquely determined by the dimensionless parameters, it is sufficient to show that two apparent shapes are not congruent when the dimensionless parameters take different values. This corresponds to verifying that the map from the dimensionless parameter set to the apparent-shape library is injective. Since it is easier to verify the contrapositive, we instead show that if two apparent shapes in the apparent-shape library are congruent, then the corresponding sets of dimensionless parameters must coincide.

The apparent shape corresponds to the tuple $\left( \sin^2 \alpha , \cos^2 \beta \right)$, where each component is a rational function. Therefore, assuming that two such tuples $\left( \sin^2 \alpha , \cos^2 \beta \right)$ are equal, it is sufficient to show that the corresponding values of the dimensionless parameters also coincide.

The strategy of the proof follows the general approach developed in Ref.~\cite{Hioki:2023ozd}. However, the key point in our argument is the irreducibility of the rational functions. A rational function is said to be irreducible if its numerator and denominator have no common factor. If two irreducible rational functions are identically equal, then their coefficients must coincide. This property does not necessarily hold when the functions are reducible. Therefore, it is essential to examine under what conditions $\sin^2 \alpha$ and $\cos^2 \beta$ are irreducible.

In Ref.~\cite{Hioki:2023ozd}, the case of Kerr black holes with two dimensionless parameters was considered. In this paper, we extend the analysis to Kerr-Newman black holes, which involve three dimensionless parameters. As a result, while the irreducibility of $\sin^2 \alpha$ and $\cos^2 \beta$ could be easily verified numerically in the Kerr case, it becomes significantly more challenging in the Kerr-Newman case.

To overcome this difficulty, we develop a new technique for verifying irreducibility. This method provides an algebraically rigorous way to determine whether the numerator and denominator of a rational function share a common factor, by computing the resultant.

The resultant of two polynomials is defined as the determinant of the Sylvester matrix, which is constructed from the coefficients of the two polynomials. If the resultant is zero, then the polynomials share a common factor.

To illustrate this concept, we explicitly compute the resultant of two quadratic (fictitious) polynomials: $3 r_{\ast} ^2 + 2 r_{\ast} + 1$ and $6 r_{\ast} ^2 + 5 r_{\ast} + 4$. In this case, the Sylvester matrix is given by:
\begin{eqnarray}
	{\rm Syl} \left( 3 r_{\ast} ^2 + 2 r_{\ast} + 1, 6 r_{\ast} ^2 + 5 r_{\ast} + 4, r_{\ast} \right) =
\begin{pmatrix}
	3 & 0 & 6 & 0 \\
	2 & 3 & 5 & 6 \\
	1 & 2 & 4 & 5 \\
	0 & 1 & 0 & 4
\end{pmatrix}
	\, .
\end{eqnarray}
Since the determinant of the Sylvester matrix yields the resultant, we compute it as follows:
\begin{eqnarray}
	{\rm Res} \left( 3 r_{\ast} ^2 + 2 r_{\ast} + 1, 6 r_{\ast} ^2 + 5 r_{\ast} + 4, r_{\ast} \right) = \det
\begin{pmatrix}
	3 & 0 & 6 & 0 \\
	2 & 3 & 5 & 6 \\
	1 & 2 & 4 & 5 \\
	0 & 1 & 0 & 4
\end{pmatrix}
	= 27 \neq 0 \, .
\end{eqnarray}
This result confirms that the two polynomials are coprime, i.e., they have no common factor.

Using this technique, we analyze the tuple $\left( \sin^2 \alpha , \cos^2 \beta \right)$ of rational functions corresponding to each element in the apparent-shape library.

The function $\sin ^2 \alpha$ is a rational function with $r_\ast$ as a variable. Note that this function is defined at $a_\ast \neq 0$. In general, a necessary and sufficient condition for two polynomials to have a common factor is that the resultant of the two polynomials is zero~\cite{cox2005}. Let us derive the condition under which the numerator and denominator of the rational function $\sin ^2 \alpha$ has a common factor:
\begin{eqnarray}
	{\rm Res} \left( p(r_\ast) , q(r_\ast) , r_\ast \right) = 0 \, ,
\end{eqnarray}
where $p(r_\ast)$ and $q(r_\ast)$ are the numerator and denominator of $\sin ^2 \alpha$, respectively. Since the resultant of the two polynomials is equivalent to the determinant of the Sylvester matrix of the two polynomials, the condition follows:
\begin{eqnarray}
	&&
	a^{20}_\ast \sin^{12}{i} \cot^8{i} \left( a^2_\ast + Q^2_\ast -1 \right)^2 \big[ 3 a^4_\ast + 8 Q^4_\ast
	\nonumber
	\\
	&&
	+8 a^2_\ast \left(2 + Q^2_\ast \right) + 4 a^2_\ast \left(4 - a^2_\ast - 2 Q^2_\ast \right) \cos{2i} + a^4_\ast \cos{4i} \big]^2 = 0 \, .
	\label{eq:resultant1}
\end{eqnarray}
Looking carefully at Eq.~(\ref{eq:resultant1}), we find that the condition can be simplified:
\begin{eqnarray}
	a^2_\ast + Q^2_\ast = 1 \;\;\; {\rm or } \;\;\; i = \frac{\pi}{2} \, .
\end{eqnarray}
This implies that the extremality of the black hole is related to the reducibility of $\sin ^2 \alpha$.

From the condition, the rational function $\sin ^2 \alpha$ is found to be irreducible in each of the four cases.
\begin{eqnarray}
	\sin ^2 \alpha
	=
	\begin{cases}
	\frac{\sum _{n=0}^{6} a_n r_\ast^n}{\sum _{n=0}^{4} b_n r_\ast^n} & \left( 0 < a^2_\ast + Q^2_\ast <1 \, {\rm and } \, i \neq \pi/2 \right) \\
	\frac{\sum _{n=0}^{4} \hat{a}_n r_\ast^n}{\sum _{n=0}^{2} \hat{b}_n r_\ast^n} & \left(  0 < a^2_\ast + Q^2_\ast <1 \, {\rm and } \, i = \pi/2 \right) \\
	\frac{\sum _{n=0}^{4} \bar{a}_n r_\ast^n}{\sum _{n=0}^{2} \bar{b}_n r_\ast^n} & \left( a^2_\ast + Q^2_\ast = 1 \, {\rm and } \, i \neq \pi/2 \right) \\
	\frac{\sum _{n=0}^{2} \tilde{a}_n r_\ast^n}{\tilde{b}_0} & \left( a^2_\ast + Q^2_\ast = 1 \, {\rm and } \, i = \pi/2 \right)
	\end{cases} ,
\label{eq:sinalpha-nonext}
\end{eqnarray}
where
\begin{eqnarray}
	a_0 &=& 4 a_\ast^4 \cos ^4 i,
	\;\;\;
	a_1 = -4 a_\ast^2 \cos ^2 i \left( a^2_\ast \cos 2i -3 a^2_\ast - 4 Q^2 \right),
	\nonumber
	\\
	a_2 &=& 9 a^4_\ast + 16 Q^4_\ast + 12 a^2_\ast \left( -1 + 2 Q^2_\ast \right) - 2 a^2_\ast \left(6 + 3 a^2_\ast + 4 Q^2_\ast \right) \cos 2i + a^4_\ast \cos^2 2i,
	\nonumber
	\\
	a_3 &=& 16 \left( -2 a^2_\ast - 3 Q^2_\ast + a^2_\ast \cos 2i \right),
	\;\;\;
	a_4 = 4 \left( 9 + 3 a^2_\ast + 4 Q^2_\ast - a^2_\ast \cos 2i \right) ,
	\nonumber
	\\
	a_5 &=& -24,
	\;\;\;
	a_6 = 4,
	\nonumber
	\\
	b_0 &=& 0,
	\;\;\;
	b_1 = 0,
	\;\;\;
	b_2 = 16 a_\ast^2 \left( a_\ast^2 + Q^2_\ast \right) \sin ^2 i,
	\;\;\;
	b_3 = -32a_\ast^2 \sin ^2 i, 
	\;\;\;
	b_4 = 16 a_\ast^2 \sin ^2 i,
	\nonumber
\end{eqnarray}
\begin{eqnarray}
	\hat{a}_0 &=& 4 \left( a_\ast^2 + Q_\ast^2 \right)^2,
	\;\;
	\hat{a}_1 = -12 \left( a_\ast^2 + Q_\ast^2 \right) ,
	\;\;
	\hat{a}_2 = 9 + 4a_\ast^2 + 4Q_\ast^2 ,
	\;\;
	\hat{a}_3 = -6,
	\;\;
	\hat{a}_4 = 1,
	\nonumber
	\\
	\hat{b}_0 &=& 4 a_\ast^2 \left( a_\ast^2 + Q_\ast^2 \right),
	\;\;\;
	\hat{b}_1 = -8a_\ast^2,
	\;\;\;
	\hat{b}_2 = 4 a_\ast^2,
	\nonumber
\end{eqnarray}
\begin{eqnarray}
	\bar{a}_0 &=& 4 a_\ast^4 \cos ^4 i,
	\;\;
	\bar{a}_1 = 16 a_\ast^2 \cos ^2 i,
	\;\;
	\bar{a}_2 = 4 \left( 4 - a_\ast^2 - a_\ast^2 \cos 2i \right) ,
	\;\;
	\bar{a}_3 = -16,
	\;\;
	\bar{a}_4 = 4,
	\nonumber
	\\
	\bar{b}_0 &=& 0,
	\;\;\;
	\bar{b}_1 = 0,
	\;\;\;
	\bar{b}_2 = 16 a_\ast^2 \sin ^2 i,
	\nonumber
\end{eqnarray}
and
\begin{eqnarray}
	\tilde{a}_0 &=& 4,
	\;\;
	\tilde{a}_1 = -4,
	\;\;
	\tilde{a}_2 = 1,
	\nonumber
	\\
	\tilde{b}_0 &=& 4a_\ast^2 \, .
	\nonumber
\end{eqnarray}

We find that $\cos ^2 \beta$ is also a rational function of $r_\ast$. As before, let us check the condition under which the numerator and denominator of this rational function $\cos ^2 \beta$ has a common factor:
\begin{eqnarray}
	{\rm Res} \left( l(r_\ast) , m(r_\ast) , r_\ast \right) = 0 \, ,
	\label{eq:resultant2-1}
\end{eqnarray}
where $l(r_\ast)$ and $m(r_\ast)$ are the numerator and denominator of $\cos ^2 \beta$, respectively. We compute the determinant of the Sylvester matrix of polynomials, $l(r_\ast)$ and $m(r_\ast)$, and write down the condition: 
\begin{eqnarray}
	r_{o\ast}^8 \left( a^2_\ast + Q^2_\ast-1 \right) ^2 \big[ a^2_\ast + Q^2_\ast + \left( r_{o\ast}-2 \right) r_{o\ast} \big] ^8 \left[ a^2_\ast + Q^2_\ast + \left( r_{o\ast} +2 \right) r_{o\ast} \right]^2 = 0 \, .
	\label{eq:resultant2-2}
\end{eqnarray}
We see that the condition for the function $\cos ^2 \beta$ to be reducible is the condition for the black hole to be extreme:
\begin{eqnarray}
	a^2_\ast + Q^2_\ast = 1 \, .
	\label{eq:resultant2-3}
\end{eqnarray}
Thus, if we divide the case into two cases, a non-extreme black hole and an extreme black hole, the function $\cos ^2 \beta$ is irreducible as follows:
\begin{eqnarray}
	\cos ^2 \beta
	=
	\begin{cases}
	\frac{\sum _{n=0}^{6} e_n r_\ast^n}{\sum _{n=0}^{6} g_n r_\ast^n} & \left(0 \leq a^2_\ast + Q^2_\ast <1 \right) \\
	\frac{\sum _{n=0}^{4} \bar{e}_n  r_\ast^n}{\sum _{n=0}^{4} \bar{g}_n r_\ast^n} & \left( a^2_\ast + Q^2_\ast = 1 \right)
	\end{cases}
	,
	\label{eq:cosbeta-nonext}
\end{eqnarray}
where
\begin{eqnarray}
	e_0 &=& r_{o\ast}^4,
	\;\;\;
	e_1 = -2 \left[ 2 \left(a_\ast^2 +Q^2_\ast \right) r_{o\ast}^2 +r_{o\ast}^4 \right],
	\;\;\;
	e_2 = r_{o\ast} \left[ 8\left(a_\ast^2 +Q^2_\ast \right) +6r_{o\ast} + r_{o\ast}^3 \right],
	\nonumber
	\\
	e_3 &=& -4\left[ \left(a_\ast^2 +Q^2_\ast \right) +4r_{o\ast} \right],
	\;\;\;
	e_4 = 9 -2 r_{o\ast} (r_{o\ast}-4),
	\;\;\;
	e_5 = -6,
	\;\;\;
	e_6 = 1,
	\nonumber
	\\
	g_0 &=& r_{o\ast}^4,
	\;\;\;
	g_1 = -2 \left[ 2\left(a_\ast^2 +Q^2_\ast \right) r_{o\ast}^2 + r_{o\ast}^4 \right],
	\;\;\;
	g_2 = 6r_{o\ast}^2 + \left[ 2\left(a_\ast^2 +Q^2_\ast \right) + r_{o\ast}^2 \right]^2,
	\nonumber
	\\
	g_3 &=& -4 \left[ 3\left(a_\ast^2 +Q^2_\ast \right) + 2r_{o\ast}^2  \right],
	\;\;\;
	g_4 = 9 + 4 \left(a_\ast^2 +Q^2_\ast \right) +2r_{o\ast}^2,
	\;\;\;
	g_5 = -6,
	\;\;\;
	g_6 = 1,
	\nonumber
\end{eqnarray}
and
\begin{eqnarray}
	\bar{e}_0 &=& r_{o\ast}^4,
	\;\;\;
	\bar{e}_1 = -4 r_{o\ast}^2,
	\;\;\;
	\bar{e}_2 = -2r_{o\ast}^2 +8r_{o\ast},
	\;\;\;
	\bar{e}_3 = -4,
	\;\;\;
	\bar{e}_4 = 1,
	\nonumber
	\\
	\bar{g}_0 &=& r_{o\ast}^4,
	\;\;\;
	\bar{g}_1 = -4r_{o\ast}^2,
	\;\;\;
	\bar{g}_2 = 4+ 2r_{o\ast}^2,
	\;\;\;
	\bar{g}_3 = -4,
	\;\;\;
	\bar{g}_4 = 1 \, .
	\nonumber
\end{eqnarray}

Let us consider the case where $a_\ast \neq 0$.

For a non-extreme Kerr-Newman black hole viewed from outside the edge-on perspective $(i \neq \pi /2)$, we assume that $\sin ^2 \alpha (r_\ast ; a_\ast, Q_\ast, i) = \sin ^2 \alpha (r_\ast ; a_\ast ^\prime, Q_\ast^\prime, i ')$. From the coefficients $b_2$ and $b_4$, we derive $a_\ast ^2 \sin ^2 i = a_\ast ^{\prime 2} \sin ^2 i^\prime$ and $a_\ast ^2 + Q_\ast^2 = a_\ast ^{\prime 2} + Q_\ast^{\prime 2}$. Analyzing the coefficient $a_3$, we conclude that $a_\ast = a_\ast ^\prime$, $Q_\ast = Q_\ast ^\prime$, and $i = i ^\prime$. Additionally, we assume that $\cos ^2 \beta (r_\ast; r_{o\ast} , a_\ast , Q_\ast ) = \cos ^2 \beta (r_\ast; r_{o\ast} ^\prime, a_\ast ^\prime , Q_\ast ^\prime )$. From the coefficient $e_0$, we deduce that $r_{o\ast} = r_{o\ast}^\prime$.

In the case of a non-extreme Kerr-Newman black hole viewed edge-on $(i = \pi/2 )$, we assume that $\sin ^2 \alpha (r_\ast ; a_\ast, Q_\ast, \pi/2) = \sin ^2 \alpha (r_\ast ; a_\ast ^\prime, Q_\ast^\prime, \pi/2)$. From the coefficients $\hat{a}_1$ and $\hat{b}_2$, we determine that $a_\ast = a_\ast ^\prime$ and $Q_\ast = Q_\ast ^\prime$. We also assume that $\cos ^2 \beta (r_\ast; r_{o\ast} , a_\ast , Q_\ast) = \cos ^2 \beta (r_\ast; r_{o\ast} ^\prime, a_\ast ^\prime , Q_\ast ^\prime)$. By comparing $e_0$, we establish that $r_{o\ast} = r_{o\ast}^\prime$.

Considering the case of an extreme Kerr-Newman black hole viewed from outside the edge-on perspective $(i \neq \pi /2)$, we assume that $\sin ^2 \alpha (r_\ast ; a_\ast, Q_\ast, i) = \sin ^2 \alpha (r_\ast ; a_\ast ^\prime, Q_\ast^\prime, i ')$. From the coefficients $\bar{a}_1$ and $\bar{b}_2$, we determine that $a = a^\prime$ and $i = i^\prime$, assuming an extreme black hole, thus $Q = Q^\prime$. We further assume that $\cos ^2 \beta (r_\ast; r_{o\ast} , a_\ast, Q_\ast) = \cos ^2 \beta (r_\ast; r_{o\ast} ^\prime, a_\ast ^\prime , Q_\ast ^\prime)$. By comparing $\bar{e}_0$, we establish that $r_{o\ast} = r_{o\ast}^\prime$.

In the case of an extreme Kerr-Newman black hole viewed edge-on $(i = \pi/2 )$, we assume that $\sin ^2 \alpha (r_\ast ; a_\ast, Q_\ast, \pi /2 ) = \sin ^2 \alpha (r_\ast ; a_\ast ^\prime, Q_\ast^\prime, \pi /2)$. By comparing $\tilde{b}_0$, we establish that $a_\ast = a_\ast^\prime$. The assumption leads us to conclude that $Q_\ast = Q_\ast^\prime$. Further assuming that $\cos ^2 \beta (r_\ast; r_{o\ast} , a_\ast, Q_\ast) = \cos ^2 \beta (r_\ast; r_{o\ast} ^\prime, a_\ast ^\prime , Q_\ast ^\prime)$, we verify that $r_{o\ast} = r_{o\ast}^\prime$ by comparing $\bar{e}_0$.

Now, let us consider the case where $a_\ast = 0$. We examine the apparent shape of the Reissner-Nordstr${\rm \ddot{o}}$m black hole. Since the Reissner-Nordstr${\rm \ddot{o}}$m black hole is spherically symmetric, it has no rotation axis, and therefore the parameter $i$ is not relevant.

The apparent shape of the Reissner-Nordstr${\rm \ddot{o}}$m black hole is a circle. As you can see from Eqs.~(\ref{eq:x}) and (\ref{eq:y}), the square of radius of the apparent shape is $4 \tan ^2 \beta /2$.

The function $\tan ^2 \beta /2$ is a monotonically decreasing function of $\cos \beta$. Note that there is a relation: $\tan ^2 \beta /2 = \left( 1 - \cos \beta \right) / \left( 1 + \cos \beta \right)$. Since $\cos \beta$ is positive, the value of $\cos ^2 \beta$ is uniquely determined for the value of $\tan ^2 \beta/2$.

The radius $r_\ast$ of the unstable spherical photon orbit of the Reissner-Nordstr${\rm \ddot{o}}$m black hole is a monotonically decreasing function of the parameter $Q_\ast$~\cite{Chandrasekhar:1985kt}:
\begin{eqnarray}
	r_\ast \left( Q_\ast  \right) = \frac{3}{2} \left[ 1 + \left( 1 - \frac{8}{9} Q^2_\ast \right) ^{1/2} \right] \, .
	\label{rofuspo}
\end{eqnarray}
We derive that there are an infinite number of tuples of $(r_{o\ast}, Q_\ast)$ for which $\cos ^2 \beta$ has the same value $\mathcal{C}$:
\begin{eqnarray}
	\cos ^2 \beta (r_\ast \left( Q_\ast \right) ; r_{o\ast} , 0, Q_\ast) = \mathcal{C} = {\rm const.}
	\label{eq:deg}
\end{eqnarray}
Using Eq.~(\ref{rofuspo}), Eq.~(\ref{eq:deg}) can be written explicitly as follows:
\begin{eqnarray}
	&& \left[ 1 + \left(1 - \frac{8}{9} Q_\ast^2 \right)^{1/2} - \frac{2}{3} r_{o\ast} \right]^2 \left( \frac{8}{3} Q_\ast^4 +  r_{o\ast} \left\{ 8 \left[ 1 + \left(1 - \frac{8}{9}Q_\ast^2\right) ^{1/2} \right]  \right.  \right.
	\nonumber
	\\
 	&& \left. + r_{o\ast} \left[ \frac{5}{3} + \left( 1 - \frac{8}{9} Q_\ast^2\right) ^{1/2} \right] \right\} - 4 Q_\ast^2 \left\{ 1 + \left( 1 - \frac{8}{9} Q_\ast^2\right) ^{1/2} \right.
	\nonumber
	\\
	&& \left. \left. + r_{o\ast} \left[ \frac{1}{3}r_{o\ast} + \frac{5}{3} + \left( 1 - \frac{8}{9} Q_\ast^2\right) ^{1/2} \right] \right\} \right) = \frac{2\mathcal{C}}{9} r_{o\ast}^4 \left[ \frac{1}{3} + \left( 1 - \frac{8}{9}Q_\ast^2 \right)^{1/2} \right]^2 \, .
	\label{eq:deg2}
\end{eqnarray}
Equation~(\ref{eq:deg2}) is a quadratic equation for $r_{o\ast}$, and by checking the coefficients, we can see that it has four real solutions. Then, for example, when $\mathcal{C}=0.9$, we can find that there are infinite number of solutions $(r_{o\ast}, Q_\ast)$ in the region $r_{o\ast} \geq 5$ for Eq.~(\ref{eq:deg2}). The apparent shapes of the Reissner-Nordstr${\rm \ddot{o}}$m black holes for such values of $(r_{o\ast}, Q_\ast)$ are all congruent. This result can be understood that when $a_\ast =0$, the value of the radius varies depending on the parameters $r_{o\ast}$ and $Q_\ast$, but it is not possible to distinguish which parameter is causing the effect.

Thus, only when the specific angular momentum is non-zero, the apparent shape on the screen coordinates of the Kerr-Newman black hole is unique. Generally, uniqueness does not hold for the apparent shape of the Kerr-Newman black hole at finite distance. The degeneracy of the shadow of the Kerr-Newman black hole at finite distance holds not only for Carter's observers, but also for general observers. This is because the Lorentz transformation for the observer acts as a conformal transformation on the observer's celestial sphere~\cite{Penrose:1985bww}. As conformal transformations are injective, the Lorentz transformation establishes a one-to-one correspondence between shadows on the celestial sphere.

\section{Parameter determination by analysis of apparent shape}
\label{sec:pc}
\subsection{Screen coordinates and Bardeen coordinates}
\label{subsec:bardeenandscreen}
We show that the parameters of the Kerr-Newman black hole can be determined from its apparent shape, along with a concrete method~\cite{Hioki:2023ozd}. For this method to be used for future shadow observations, we assume that $r_{o}$ is sufficiently large. This means that we consider the apparent shape on the so-called Bardeen coordinates $(b_x , b_y)$ for the observer at spatial infinity. Note that the apparent shape on the screen coordinates of the Kerr-Newman black hole for an observer at spatial infinity is a point. The Bardeen coordinates $(b_x , b_y)$ is defined as follows:
\begin{eqnarray}
	b_x &\coloneqq& \lim_{r \to \infty} \frac{-r k^{(\phi)}}{k^{(t)}} = \left( \ell + \ell _p \right) \csc i \, , \\
	b_y &\coloneqq& \lim_{r \to \infty} \frac{r k^{(\theta)}}{k^{(t)}} = \left( \mathbb{Q} + a^2 \cos ^2 i - \ell \cot ^2 i \right)^{1/2} \, ,
\end{eqnarray}
where $\left( k^{(t)}, k^{(r)}, k^{(\theta)} , k^{(\phi)} \right)$ are the tetrad components of the four-momentum of the massless test particle.

Although $b_x$ contains a term for the conserved quantity $\ell _p$ of the principal null geodesic, this term is not present in the Bardeen coordinates for a zero-angular-momentum observer~\cite{EventHorizonTelescope:2021dqv}.
Since we are now assuming a Carter's observer, the origin of the Bardeen coordinates is shifted~\cite{Perlick:2021aok}. The analysis in this paper does not use information about the position on the Bardeen coordinates of the apparent shape. Therefore, the results of the analysis do not change for our assumed observer or for the case of the zero-angular-momentum observer.

Let us now describe the transformation from the screen coordinates to the Bardeen coordinates. In brief, Bardeen coordinates are the linearization of the screen coordinates of the distant observer~\cite{Perlick:2021aok, Tsukamoto:2024gkz}, as follows:
\begin{eqnarray}
	x &=& b_x r_{o}^{-1} + O \left( r_{o}^{-2}  \right) \,\; \left( r_{o} \rightarrow \infty \right) \, , \\
	y &=& b_y r_{o}^{-1} + O \left( r_{o}^{-2}  \right) \,\; \left( r_{o} \rightarrow \infty \right) \, .
\end{eqnarray}
From here on in the following subsections in Sec.~\ref{sec:pc}, the apparent shape on the Bardeen coordinates of the Kerr-Newman black hole will be the target of our analysis.

\subsection{Uniqueness of apparent shape on the Bardeen coordinates}
\label{sec:uniq2}
We prove the uniqueness of the apparent shape on the Bardeen coordinates of the Kerr-Newman black hole. Even if the apparent shape on the screen coordinates of a Kerr-Newman black hole is not unique, it does not necessarily mean that the apparent shape on the Bardeen coordinates of the Kerr-Newman black hole is non-unique.

The functions $b_x$ and $b_y^2$ are rational functions with $r_\ast$ as a variable. Note that the function is defined at $a_\ast \neq 0$. First, we analyze the rational function $b_x$. The condition under which the resultant for the numerator and denominator of the rational function $b_x$ vanishies is as follows:
\begin{eqnarray}
	a_\ast^3 \left( a_\ast^2 + Q_\ast^2 -1 \right) \sin ^3 i = 0 \, .
	\label{eq:resultant3}
\end{eqnarray}
The irreducible rational function $b_x$ follows:
\begin{eqnarray}
	b_x
	=
	\begin{cases}
	\frac{\sum _{n=0}^{3} s_n r_\ast^n}{\sum _{n=0}^{1} t_n r_\ast^n} & \left(0 < a^2_\ast + Q^2_\ast <1 \right) \\
	\frac{\sum _{n=0}^{2} \bar{s}_n  r_\ast^n}{\bar{t}_0} & \left( a^2_\ast + Q^2_\ast = 1 \right)
	\end{cases}
	,
	\label{eq:bx}
\end{eqnarray}
where
\begin{eqnarray}
	s_0 &=& 2 a_\ast^2 \cos ^2 i,
	\;\;\;
	s_1 = 3 a_\ast^2 + 4 Q_\ast ^2 - a_\ast^2 \cos 2i,
	\;\;\;
	s_2 = -6,
	\;\;\;
	s_3 = 2,
	\nonumber
	\\
	t_0 &=& -2 a_\ast \sin i,
	\;\;\;
	t_1 = 2 a_\ast \sin i,
	\nonumber
\end{eqnarray}
and
\begin{eqnarray}
	\bar{s}_0 &=& -2 a_\ast^2 \cos ^2 i,
	\;\;\;
	\bar{s}_1 = -4,
	\;\;\;
	\bar{s}_2 = -2,
	\nonumber
	\\
	\bar{t}_0 &=& 2 a_\ast \sin i \, .
	\nonumber
\end{eqnarray}

Next, the condition under which the resultant for the numerator and denominator of the rational function $b_y^2$ is zero is as follows:
\begin{eqnarray}
	a_\ast^{12} \left( a_\ast^2 + Q_\ast^2 -1 \right) ^2 \left( a_\ast^2 \cos ^2 i + Q_\ast^2 -1  \right) ^2 \sin ^{12} i = 0 \, .
	\label{eq:resultant4}
\end{eqnarray}
Then the irreducible rational function $b_y^2$ is as follows:
\begin{eqnarray}
	b^2_y
	=
	\begin{cases}
	\frac{\sum _{n=0}^{6} u_n r_\ast^n}{\sum _{n=0}^{2} v_n r_\ast^n} & \left(0 < a^2_\ast + Q^2_\ast <1 \right) \\
	\frac{\sum _{n=0}^{4} \bar{u}_n  r_\ast^n}{\bar{v}_0} & \left( a^2_\ast + Q^2_\ast = 1 \right)
	\end{cases}
	,
	\label{eq:by}
\end{eqnarray}
where
\begin{eqnarray}
	u_0 &=& -8 a_\ast^4 \cos ^4 i,
	\;\;\;
	u_1 = 8 a_\ast^2 \cos ^2 i \left( -3 a_\ast^2 -4 Q_\ast^2 + a_\ast^2 \cos 2i \right),
	\nonumber
	\\
	u_2 &=& -3 a^4 -32 Q_\ast^4 + 8 a_\ast^2 \left( 3 - 4 Q_\ast^2 \right) -4a_\ast^2 \left(  -6 +a_\ast^2 \right) \cos 2i - a^4 \cos 4i ,
	\nonumber
	\\
	u_3 &=& 32 \left( a_\ast^2 +3 Q_\ast^2 \right),
	\;\;\;
	u_4 = -8 \left( 9 + a_\ast^2 + 4Q_\ast^2 + a_\ast^2 \cos 2i \right),
	\;\;\;
	u_5 = 48,
	\;\;\;
	u_6 = -8,
	\nonumber
	\\
	v_0 &=& 8 a_\ast^2 \sin ^2 i,
	\;\;\;
	v_1 = -16 a_\ast^2 \sin^2 i,
	\;\;\;
	v_2 = 8 a_\ast^2 \sin^2 i,
	\nonumber
\end{eqnarray}
and
\begin{eqnarray}
	\bar{u}_0 &=& -8a_\ast^4 \cos ^3 i \cot i,
	\;\;\;
	\bar{u}_1 = -32 a_\ast^2 \cos i \cot i,
	\;\;\;
	\bar{u}_2 = -16 \left( a_\ast^2 -2  \right) \csc i  +16 a_\ast^2 \sin i,
	\nonumber
	\\
	\bar{u}_3 &=& 32 \csc i,
	\;\;\;
	\bar{u}_4 = - 8 \csc i,
	\nonumber
	\\
	\bar{v}_0 &=& 8 a_\ast^2 \sin i.
	\nonumber
\end{eqnarray}

Let us consider the case where $a_\ast \neq 0$.

For a non-extreme Kerr-Newman black hole, we assume that $b_x (r_\ast ; a_\ast, Q_\ast, i) = b_x (r_\ast ; a_\ast ^\prime, Q_\ast^\prime, i^\prime )$. From the coefficient $t_1$, we derive $a_\ast \sin i = a_\ast ^{\prime } \sin i^\prime$. Analyzing the coefficient $s_0$ and $s_1$, we conclude that $a_\ast = a_\ast ^\prime$ and $i = i ^\prime$. Additionally, we assume that $b_y^2 (r_\ast; a_\ast , Q_\ast , i ) = b_y^2 (r_\ast; a_\ast ^\prime , Q_\ast ^\prime, i^\prime )$. From the coefficient $u_3$, we deduce that $Q = Q^\prime$.

Considering the case of an extreme Kerr-Newman black hole, we assume that $b_x (r_\ast ; a_\ast, Q_\ast, i) = b_x (r_\ast ; a_\ast ^\prime, Q_\ast^\prime, i^\prime )$. From the coefficient $\bar{t}_0$, we determine that $a_\ast \sin i = a_\ast ^{\prime } \sin i^\prime$. Furthermore, we assume that $b_y^2 (r_\ast; a_\ast, Q_\ast, i) = b_y^2 (r_\ast ; a_\ast ^\prime, Q_\ast^\prime, i^\prime )$. By comparing $\bar{u}_3$, we establish that $i = i^\prime$, which implies $a_\ast = a_\ast^\prime$. Assuming an extreme black hole, it follows that $Q = Q^\prime$.

Now we consider the case where $a_\ast =0$.

In this case, we examine the apparent shape of the Reissner-Nordstr${\rm \ddot{o}}$m black hole. Since the Reissner-Nordstr${\rm \ddot{o}}$m black hole is spherically symmetric, there is no rotation axis, and thus, the parameter $i$ is not needed.

The apparent shape of the Reissner-Nordstr${\rm \ddot{o}}$m black hole is a circle. The radius of the circle is a monotonically decreasing function with respect to the parameter $Q_\ast$~\cite{Zakharov:2005ek}. This means that the apparent shape of the Reissner-Nordstr${\rm \ddot{o}}$m black hole is unique.

Thus, the apparent shape of the Kerr-Newman black hole on the Bardeen coordinates is unique.

\subsection{Map from parameter space to apparent-shape library}
As we saw in the previous section, for a given parameters, an apparent shape can be determined. From this, we can consider a map from a parameter space to an apparent-shape library. The parameter space $P$ is defined as
\begin{eqnarray}
	P &\coloneqq& \left\{ \left( M, a, Q, i \right) \biggm\vert 0 \neq \frac{a}{M}, 0 \leq a^2 + Q^2 \leq M^2, 0 < i \leq \frac{\pi}{2} \right\} \cup P_0  \, ,
	\label{eq:pspace}
\end{eqnarray}
where
\begin{eqnarray}
	P_0 \coloneqq \left\{ \left( M, 0, Q, 0 \right) \biggm\vert 0 \leq \frac{Q}{M} \leq 1 \right\} \, .
\end{eqnarray}
When the specific angular momentum $a$ is zero, the black hole is spherically symmetric and has no axis of rotation. Hence, the inclination angle $i$ need not be considered. The subset $P_0$ of $P$ corresponds to the case where the black hole is a Reissner-Nordstr${\rm \ddot{o}}$m black hole. The apparent-shape library $S$ is defined as the set of all possible apparent shapes that can be generated by the system of the black hole, light source, and observer. The map $f_1$ from the parameter space $P$ to the apparent-shape library $S$ is defined as
\begin{eqnarray}
	f_1 : P \ni \left( M, a, Q, i \right) \mapsto \{ \left. \left( b_x(r_s) , b_y(r_s) \right) \right| r_s \in I \} \in S \, .
\end{eqnarray}

The apparent shape is not explicitly dependent on mass $M$, but is determined by four dimensionless parameters $\left( r_{o\ast} , a_\ast , Q_\ast , i \right)$. Then, we define the equivalence relation $\sim$ in parameter space $P$ as follows:
\begin{eqnarray}
	\left( M, a, Q, i \right) \sim \left( M', a', Q', i' \right)
	\stackrel{\rm def.}{\iff}
	\begin{cases}
		a/M = a'/M' \\
		Q/M = Q'/M' \\
		i = i' 
	\end{cases} .
\end{eqnarray}
In Sec.~\ref{sec:pc}, we see that the following map $\tilde{f}_1$ is injective:
\begin{eqnarray}
	\tilde{f}_1: P/\sim \ni [(M, a, Q, i)] \mapsto f_1((M, a, Q, i)) \in S \, .
\end{eqnarray}
That is, each equivalence class $\left[  \left( M, a, Q, i \right) \right]$ in quotient space $P/\sim$ is mapped to one element in the apparent-shape library $S$. This is a paraphrase of the proposition that an apparent shape of Kerr-Newman balck hole on Bardeen coordinates is unique.

\subsection{Observables of apparent shape}
\label{sec:observables}
\begin{figure}[tb]
		\begin{tabular}{ ccc }
			\includegraphics[height=4.4cm]{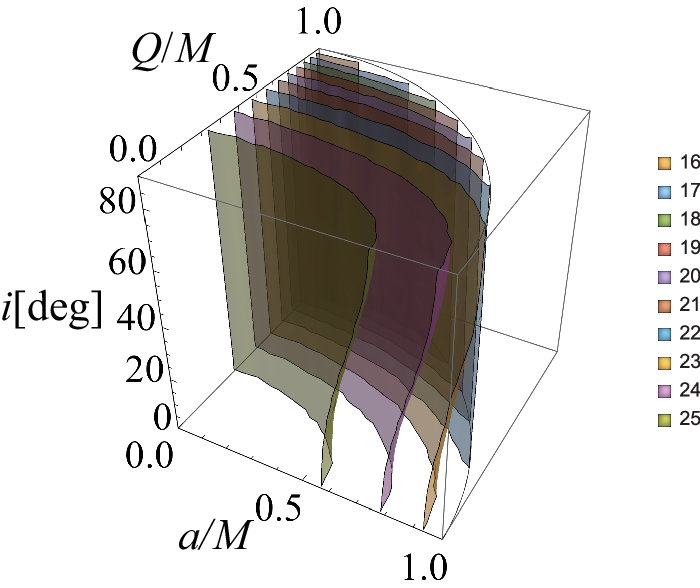} &
			\includegraphics[height=4.4cm]{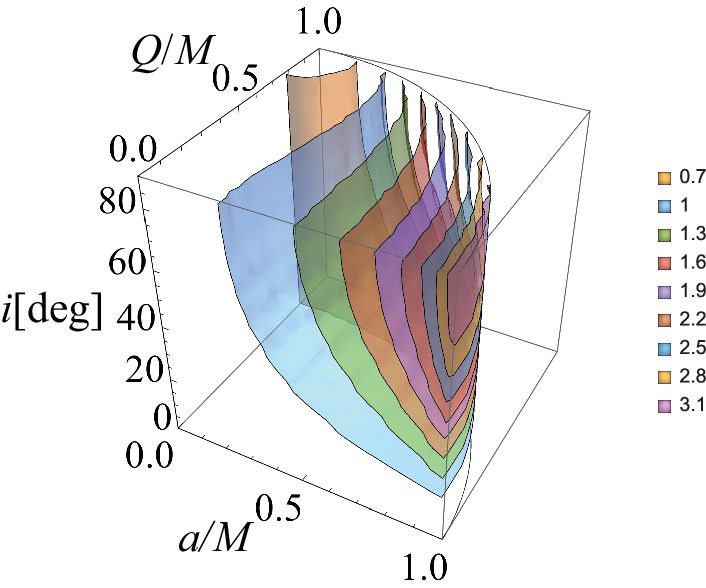} &
			\includegraphics[height=4.4cm]{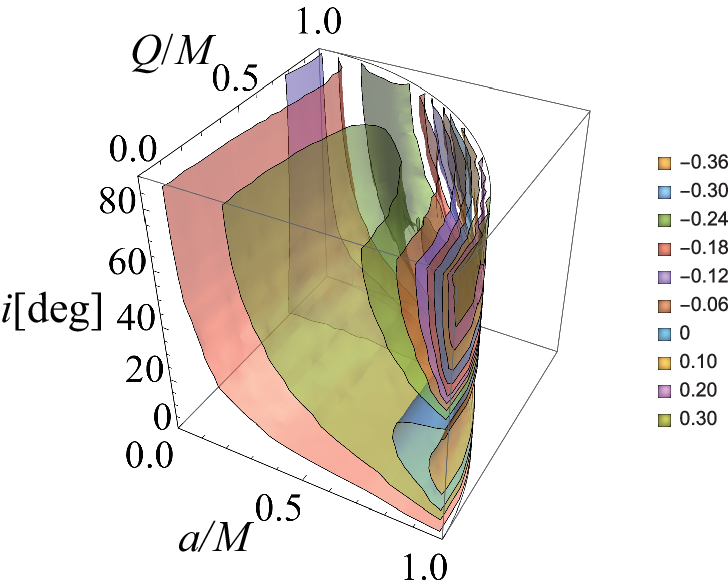} \\
			(a) $z_1$ & (b) $z_2$ & (c) $z_3$ \\
		\end{tabular}
	\caption{\footnotesize{(a) An isosurface map of the size $z_1$. (b) An isosurface map of the primary distortion $z_2$. (c) An isosurface map of the secondary distortion $z_3$.}}
	\label{fig050}
\end{figure}
We propose the use of observables to distinguish between different apparent shapes. An observable is defined as a measurable quantity derived from an apparent shape that captures its key features. To characterize the apparent shape, we introduce three key observables: size $z_1$, primary distortion $z_2$, and secondary distortion $z_3$. These observables are constructed from the Fourier coefficients of the apparent shape, with the detailed procedure described in Appendix~\ref{sec:features}.

By applying an orthogonal transformation to the Fourier coefficients up to the eleventh order, we obtain eleven quantities. From these, we select three that primarily characterize the apparent shape. The orthogonal transformation is performed using principal component analysis (PCA), a standard technique for dimensionality reduction in multivariable data~\cite{Jolliffe, Hioki:2023ozd}. The transformed quantities are referred to as principal components. We provide the transformation matrix $\bm{A}$ of the orthogonal transformation as Supplemental Material to the paper~\cite{SupplmentalMaterial1}.

The isosurfaces of the observables are depicted in Fig.~\ref{fig050}. We have provided data on the parameters and their corresponding principal components in Supplemental Material to the paper~\cite{SupplmentalMaterial2}.

In the next subsection, we show that the map $f$ from quotient parameter space $P/\sim$ to observables space $O$ is injection, where we define the set of all values taken by observables $(z_1, z_2, z_3)$ to be observable space $O$. We then conclude that the dimensionless parameters $(a_\ast , Q_\ast , i)$ can be determined from the apparent shape.

\subsection{Parameter determination}
\label{sec:determination}
\begin{figure}[tb]
		\begin{tabular}{ ccc }
			\includegraphics[height=4.3cm]{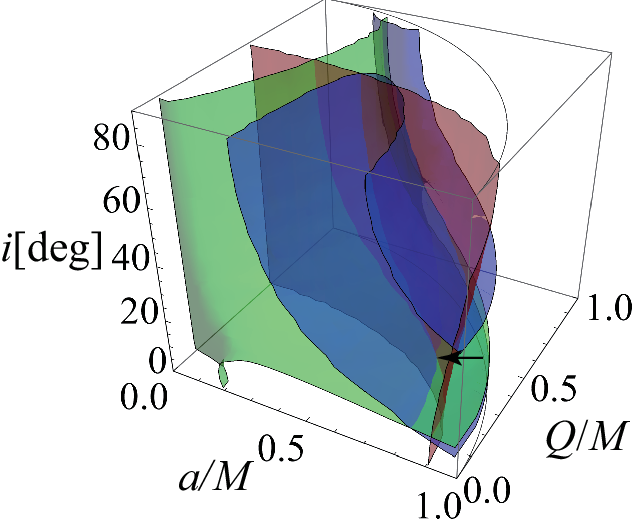} &
			\includegraphics[height=4.3cm]{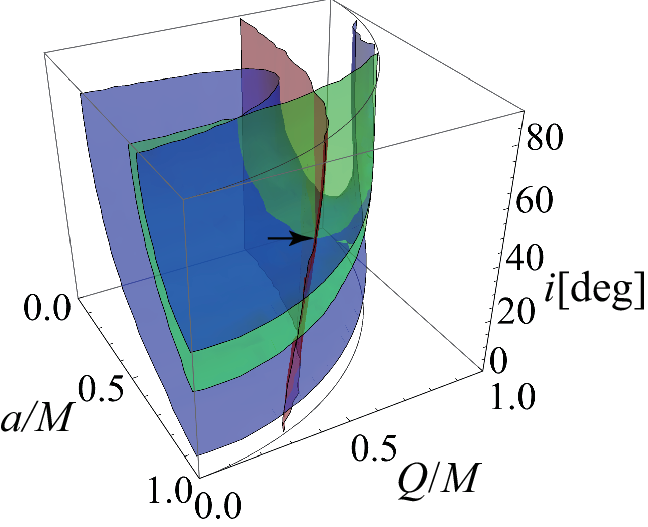} &
			\includegraphics[height=4.3cm]{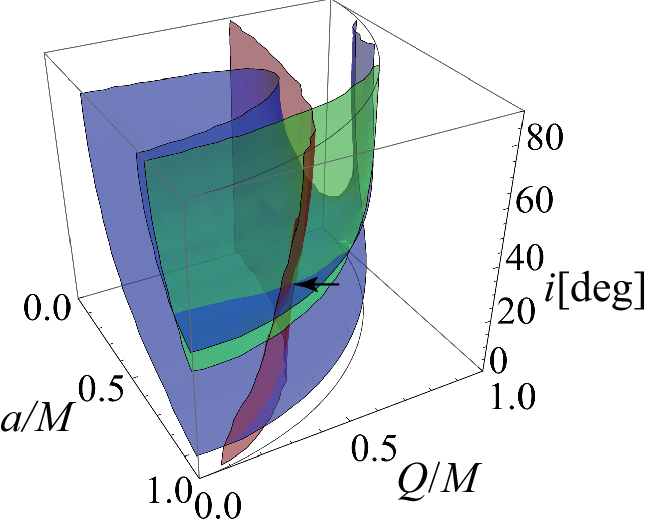} \\
			(a) $z_2=0.9$, $z_3=-0.2456$ & (b) $z_2=1.6$, $z_3=-0.2456$ & (c) $z_2=1.9$, $z_3=-0.2456$ \\
			\includegraphics[height=4.3cm]{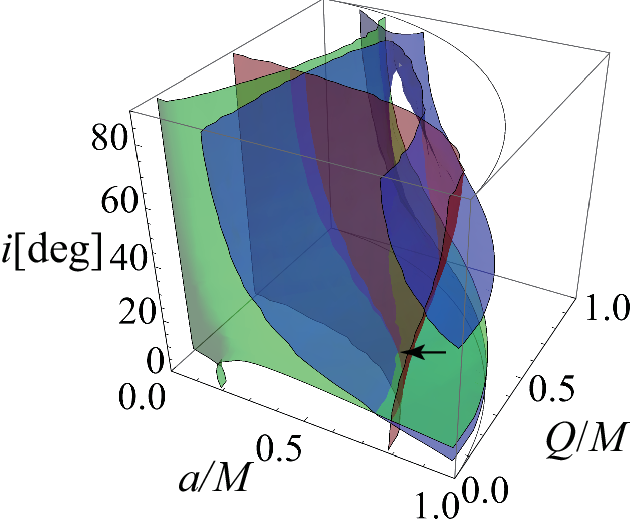} &
			\includegraphics[height=4.3cm]{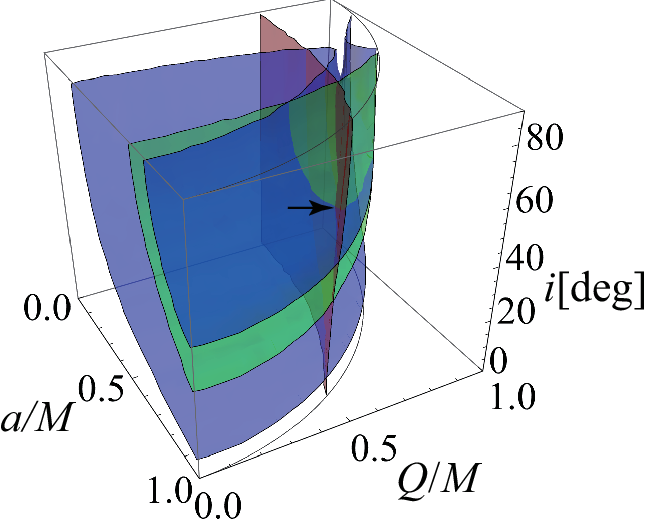} &
			\includegraphics[height=4.3cm]{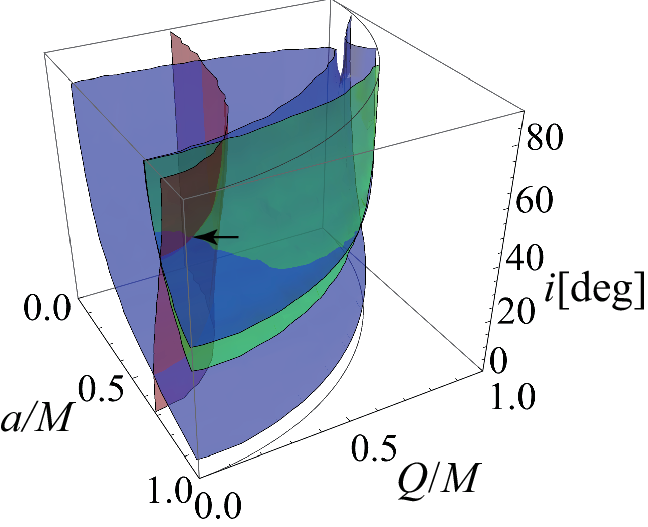} \\
			(d) $z_2=0.9$, $z_3=-0.225$ & (e) $z_2=1.6$, $z_3=-0.225$ & (f) $z_2=1.9$, $z_3=-0.225$ \\
			\includegraphics[height=4.3cm]{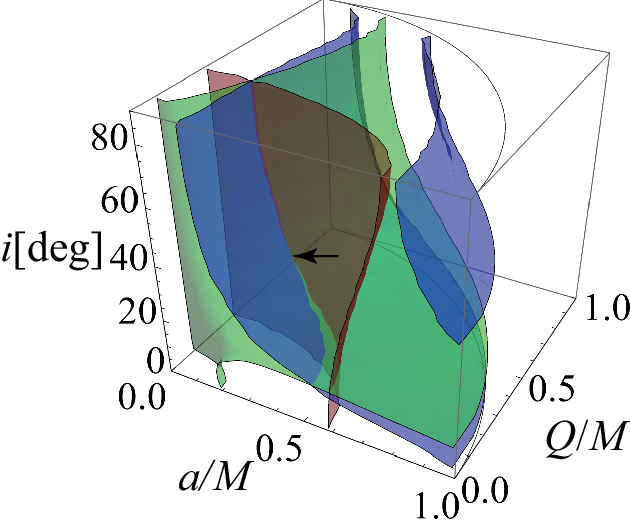} &
			\includegraphics[height=4.3cm]{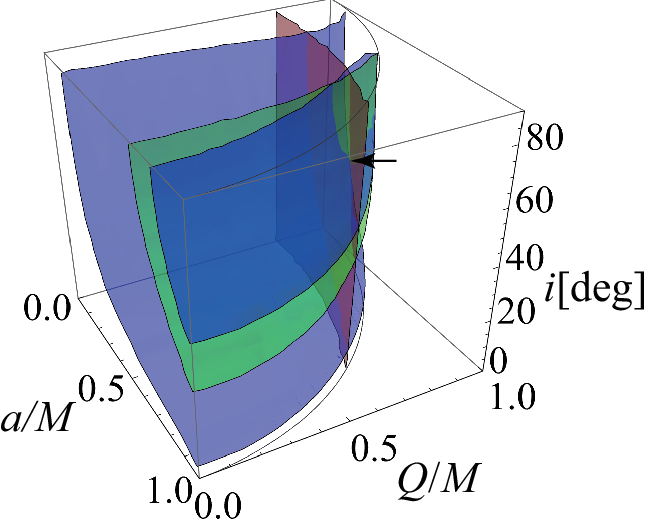} &
			\includegraphics[height=4.3cm]{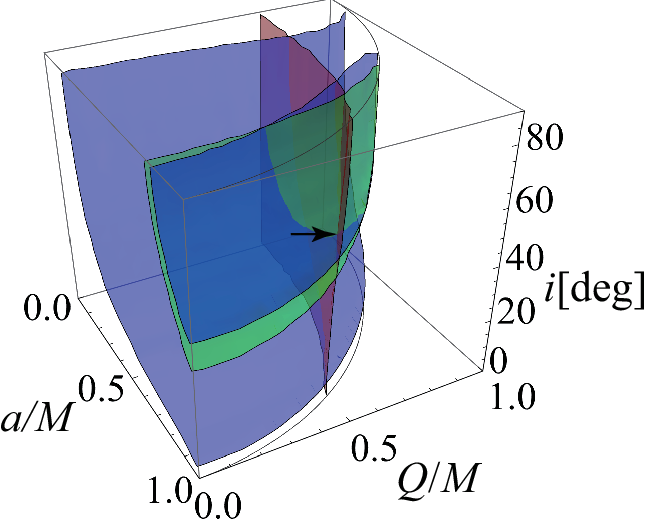} \\
			(g) $z_2=0.9$, $z_3=0.2$ & (h) $z_2=1.6$, $z_3=0.2$ & (i) $z_2=1.9$, $z_3=0.2$ \\
		\end{tabular}
	\caption{\footnotesize{The isosurfaces of the size $z_1$ (the red sufaces), the primary distortion $z_2$ (the green surfaces), and the secondary distortion $z_3$ (the blue surfaces). In (a), (b), (c), (d), (e), (f), (g), (h), and (i), sizes $z_1$ are 23.2, 22, 22.5, 24, 21, 24.5, 24.95, 20, and, 21, respectively.}}
	\label{fig090}
\end{figure}
Figure~\ref{fig050}(a) shows that the size of the apparent shape is almost independent of the inclination angle $i$. When the specific angular momentum is zero, the intersection of the isosurfaces with the $Q_\ast$-$i$ plane is parallel to the $i$-axis. This is because when $a_\ast = 0$, the black hole has no axis of rotation and its apparent shape is a circle. Additionally, we can say that the larger the electric charge $Q_\ast$, the smaller the size. As shown in Fig.~\ref{fig050}(b), the primary distortion is the deformation from the circle of the apparent shape, which differs between regions of small and large specific angular momentum $a_\ast$. The secondary distortion, depicted in Fig.~\ref{fig050}(c), represents the common distortion effect between regions of small and large specific angular momentum $a_\ast$.

Varying the values of the three observables $(z_1, z_2, z_3)$, the isosurfaces of the three observables are drawn in Fig.~\ref{fig090}. As can be seen in Fig.~\ref{fig090}, in all cases, the isosurfaces of the three observables intersect at a single point. This indicates that the map $f$ from $P/\sim$ to $O$ is injective. In other words, if the apparent shape of the black hole can be obtained by observation, the dimensionless parameters $(a_\ast, Q_\ast, i)$ can be uniquely determined. Eventually, we can determine all four parameters $(M, a, Q, i)$ if even one value of $M$, $a$ or $Q$ is known a priori from another observation.

\section{Discussion}
\label{sec:discussion}
The shadow of the Kerr-Newman black hole has been extensively studied~\cite{Takahashi:2005hy, Kraniotis:2014paa, Tsukamoto:2017fxq, Hsiao:2019ohy, Xavier:2020egv, Wang:2022ouq, Guo:2024mij}. In particular, the theme of Ref.~\cite{Takahashi:2005hy} is closely related to the theme of this paper. Reference~\cite{Takahashi:2005hy} gives a concrete example of the apparent shape of a Kerr-Newman black hole when observing from spatial infinity. In addition, it proposes a shadow feature to measure the dimensionless parameter $(a/M, Q/M, i)$ of the Kerr-Newman black hole.

Although the proposal of shadow features in Ref.~\cite{Takahashi:2005hy} is pioneering, they do not show whether the dimensionless parameters of a Kerr-Newman black hole can really be measured from the shadow features. On the other hand, our paper demonstrates, both analytically and numerically, that dimensionless parameters can be measured by using systematically designed shadow features.

Furthermore, the shadow features proposed in Ref.~\cite{Takahashi:2005hy} require information on the position of the singularity on the screen coordinates. This implicitly assumes the implementation of another observation. On the other hand, we show that dimensionless parameters can be measured by methods using shadow features based only on shadow size and shape.

The setup of the system consisting of the black hole, the source, and the observer is described in Sec.~\ref{sec:setup}, but let us summarize the model assumptions here as follows. (1) Assume the existence of a Kerr-Newman black hole without an accretion disk as a candidate black hole to be observed. (2) As the observer, we assume a Carter's observer. (3) The light source is uniformly distributed on a sphere with a radius greater than the distance between the black hole and the observer, emitting light isotropically.

The model considered in this paper is simple, yet highly relevant to realistic black hole scenarios. The shadow contour of the black hole in this model is referred to as the critical curve. It consists of null geodesics that wind around an unstable spherical photon orbit an infinite number of times before reaching the observer.

In images of black holes with accretion disks, there appears a feature known as the photon ring. It is composed of null geodesics that orbit the unstable spherical photon orbit one or more times before reaching the observer. The critical curve corresponds to the asymptotic inner boundary of the photon ring, approached as the number of windings increases.

Although the model adopted here is simplified to emphasize the essential nature of the critical curve, the primary aim of the analysis is to clarify the properties of the photon ring, a defining feature in realistic images of black holes surrounded by accretion disks.

There have been many attempts to put a constraint on the magnitude of the charge using critical curve~\cite{Tsukamoto:2014tja, Kuang:2022ojj, Gomez:2024ack, Tsukamoto:2024gkz}. A comprehensive study that places a constraint on the black hole charges from critical curve conducted by the EHT collaboration~\cite{EventHorizonTelescope:2021dqv} would be helpful in this area.

The method proposed in this paper for extracting black hole parameters from shadow observations is broadly applicable and can be extended to a variety of black hole models beyond the one considered here.

For instance, in the case of a black hole with a thin accretion disk, the method can be applied to the shape of the inner edge of the direct image of the disk, as well as to the photon ring itself. Furthermore, when the light source is modeled using GRMHD simulations, a time series of black hole images (a movie) can be produced. In the time-averaged, static image derived from such a simulation, this method can also be applied to contour lines at half the peak brightness.

Specifically, using this method, one can extract characteristic curves from images of various black hole models, compute their Fourier coefficients, and analyze the principal components of these coefficients. This makes it possible to assess whether black hole parameters can be reliably inferred from shadow observations.

As a first step, we consider a simplified model of a black hole without an accretion disk or accretion flow. Even for highly idealized models, establishing whether dimensionless parameters can be determined from shadow observations is challenging, as it requires constructing a robust validation methodology from the ground up.

Carter's observer becomes a static observer at spatial infinity. For this reason, the assumption of the existence of Carter's observers is not special, but natural. Moreover, the degeneracy of the shadow of the Kerr-Newman black hole at finite distance holds not only for Carter's observers, but also for general observers. This is because the Lorentz transformation for the observer acts as a conformal transformation for the observer's celestial sphere~\cite{Penrose:1985bww}. Since conformal transformation is injective, the Lorentz transformation is a one-to-one correspondence from one shadow to another on the celestial sphere.

Note that our goal is not to test the Einstein-Maxwell theory. We are proving whether it is possible to determine the dimensionless parameters of a black hole from the size and shape of its shadow, assuming the Einstein-Maxwell theory.

We clarify the differences between the results of this paper and those of Ref.~\cite{Medeiros:2019cde}. Reference~\cite{Medeiros:2019cde} demonstrates that the shadow of a non-Kerr black hole with more than three dimensionless parameters can be reconstructed using a small number of principal components. However, it does not verify whether the shadows are degenerate with respect to these parameters.

In this paper, we investigate whether such degeneracy exists and show that the shadows of Kerr-Newman black holes are non-degenerate when observed from spatial infinity. Our analysis focuses on determining whether the dimensionless parameters can be inferred from shadow observations, distinguishing it from the approach taken in Ref.~\cite{Medeiros:2019cde}. In particular, we prove that the dimensionless parameters of a Kerr-Newman black hole can be uniquely determined from its shadow observed at spatial infinity, a result not verified in Ref.~\cite{Medeiros:2019cde}.

We have only outlined in Sec.~\ref{sec:observables} the procedure for constructing shadow features. The details of the procedure are given in Appendix~\ref{sec:features}. The following overview of how to obtain shadow features is sufficient to understand the results of this paper. We propose observables of apparent shape to distinguish between all shadows. An observable is defined as a quantity that can be derived from a critical curve to describe its features. We utilize three observables to characterize the critical curve: size $z_1$, primary distortion $z_2$, and secondary distortion $z_3$. These observables are derived from the Fourier coefficients of the critical curve. By orthogonal transformation of the Fourier coefficients up to the eleventh, we obtain eleven quantities. We select three quantities that mainly characterize the critical curve. This orthogonal transformation is conducted using principal component analysis, a standard data analysis method that reduces dimensions in multivariable data~\cite{Jolliffe, Hioki:2023ozd}. The orthogonally transformed quantities are referred to as principal components.

Let us explain why we need eleven Fourier coefficients. First, there was no one-to-one correspondence between the three dimensionless parameters and three Fourier coefficients. Therefore, we increased the Fourier coefficients one by one, and for the first time, a one-to-one correspondence was confirmed between the three dimensionless parameters and three principal components using eleven Fourier coefficients. This is the minimum number of Fourier coefficients needed to determine the dimensionless parameters from the size and shape of the critical curve.

In this paper, we examined whether the map $f$ from the three dimensionless parameters of the black hole to the three features of the critical curve is a one-to-one correspondence. In Fig.~\ref{fig050}, we illustrate the variation of the three features of the critical curve with respect to the dimensionless parameters by plotting their isosurfaces. In Fig.~\ref{fig090}, it can be confirmed that the three isosurfaces intersect at a single point. This shows that the map $f$ is a one-to-one correspondence. In Fig.~\ref{fig040}, we also show a conceptual diagram of the map $f$ to be verified, which is explained in detail in Appendix~\ref{sec:features}.

We reiterate that many studies have been conducted to constrain the physical parameters of black holes using observational data reported by the EHT collaboration~\cite{Tsukamoto:2014tja, EventHorizonTelescope:2021dqv, Kuang:2022ojj, Gomez:2024ack, Tsukamoto:2024gkz}. These studies are based on methods for constraining dimensionless parameters from the shape of the black hole shadow~\cite{Hioki:2009na}. The conventional approach is applicable to systems with two dimensionless parameters. Even if the uniqueness of the shadow associated with an assumed black hole solution is not verified, the method can still constrain one of the two parameters by assuming a fixed value for the other. If the uniqueness of the shadow can be established within the assumed theory, both dimensionless parameters can be determined simultaneously.

The method proposed in this paper is applicable to systems with three or more dimensionless parameters. In such cases, it is generally difficult to intuitively identify the quantities that characterize the shadow, as in the two-parameter case. However, our method enables the systematic construction of such quantities. If observational data include not only the size but also the distortion of the shadow, this method can be used to constrain all three dimensionless parameters from the data. Furthermore, we analytically prove the uniqueness of the shadow of a Kerr-Newman black hole observed from spatial infinity. Therefore, under the assumption that the observed black hole is the Kerr-Newman black hole, all three dimensionless parameters can be determined simultaneously from the observational data.

Thus, this paper extends the method for determining black hole parameters from observational data, making it applicable to a broader class of black hole solutions. In the future, establishing the uniqueness of black hole shadows in various theoretical models will lay a solid foundation for the development of a general theory of black hole imaging and shadow analysis.

We also discuss some fundamental approaches to the theoretical study of black hole shadows. There are two main approaches: mathematical and numerical. Each has distinct strengths and limitations and should be chosen appropriately based on the configuration of the black hole, light source, and observer.

The mathematical approach offers the advantage of yielding rigorous results. When an analytical solution exists, it can provide precise descriptions of the phenomena. Additionally, transforming the equations analytically can lead to a deeper understanding of the properties of the black hole shadow. However, when modeling realistic light sources, it is often difficult to obtain analytical results.

On the other hand, the numerical approach computes black hole images via simulations. Its strength lies in its ability to handle complex light sources that cannot be addressed analytically, making it indispensable in black hole shadow studies. However, it is subject to numerical errors, such as rounding and discretization, which must be carefully evaluated. Moreover, numerical results may lack direct physical insight and can be difficult to interpret when treated as a black box.

In conclusion, the mathematical and numerical approaches each play an essential role. A complete understanding of black hole shadows requires both. The mathematical approach ensures theoretical rigor, while the numerical approach allows for the study of realistic scenarios. Together, they form the foundation for advancing black hole shadow research.

\section{Conclusion}
\label{sec:conclusion}
We have analyzed the apparent shape of Kerr-Newman black holes without accretion disks, i.e., bare Kerr-Newman black holes. Our investigation focused on determining dimensionless parameters from the apparent shape, considering cases where the observer is positioned at either a finite distance or spatial infinity. The dimensionless parameters of the system, comprising the black hole, light source, and observer, are the dimensionless specific angular momentum $a_\ast$, the dimensionless electric charge $Q_\ast$, the inclination angle $i$, and the dimensionless distance $r_{o\ast}$

We analytically proved that the apparent shape of a Kerr-Newman black hole in Bardeen coordinates is unique. A crucial point in this proof was that the equation governing the apparent shape is an irreducible rational function. This uniqueness, in principle, implies that the dimensionless parameters $(a_\ast, Q_\ast, i)$ of the Kerr-Newman black hole can be determined from its apparent shape.

Additionally, we demonstrated that these dimensionless parameters $(a_\ast, Q_\ast, i)$ can be extracted from the apparent shape in Bardeen coordinates using our proposed method. The key to this method is that the three principal components $(z_1, z_2, z_3)$ of the eleven Fourier coefficients of the apparent shape in Bardeen coordinates correspond one-to-one with the three dimensionless parameters $(a_\ast, Q_\ast, i)$. Our approach could be applied in the future to determine parameters for various black hole solutions based on the apparent shape, particularly when both the size and distortion of the black hole shadow are observable.

Interestingly, we also analytically showed that the apparent shape of a Kerr-Newman black hole at a finite distance is not unique. For instance, the apparent shape of a Kerr-Newman black hole with zero specific angular momentum is a circle. Although both the distance $r_{o\ast}$ and the electric charge $Q_\ast$ influence the radius of this circle, it was found that these two parameters cannot be independently determined from the radius, even if the apparent shape is observed.

The uniqueness of the apparent shapes in Bardeen coordinates, contrasted with the non-uniqueness in screen coordinates, reveals an aspect of the structure of the image library.

In this paper, we simplified our model by assuming a black hole without an accretion disk. The next step will involve analyzing a more realistic black hole model that includes an accretion disk. In that case, the uniqueness of the apparent shape of the black hole at finite distance may hold. Our results strengthen the direction of research in black hole shadow observation.

\section*{Acknowledgements}
This work was supported in part by JSPS KAKENHI Grant No. JP22K03623 (U.M.).

\appendix

\section{Procedure for constructing shadow observables}
\label{sec:features}
\begin{figure}[tb]
		\begin{tabular}{ c }
			\includegraphics[height=3.8cm]
            {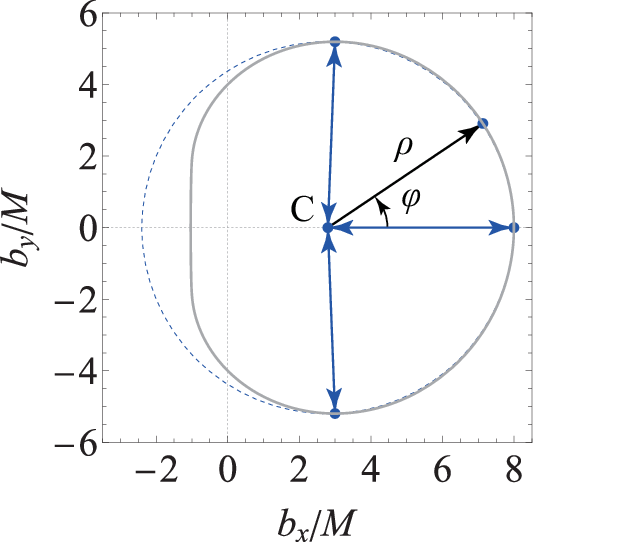}
		\end{tabular}
	\caption{\footnotesize{Polar coordinates $(\rho ,\varphi)$ with origin at the center C of the apparent shape of the Kerr-Newman black hole.}}
	\label{fig030}
\end{figure}
\begin{figure}[tb]
		\begin{tabular}{ c }
			\includegraphics[height=5.5cm]
            {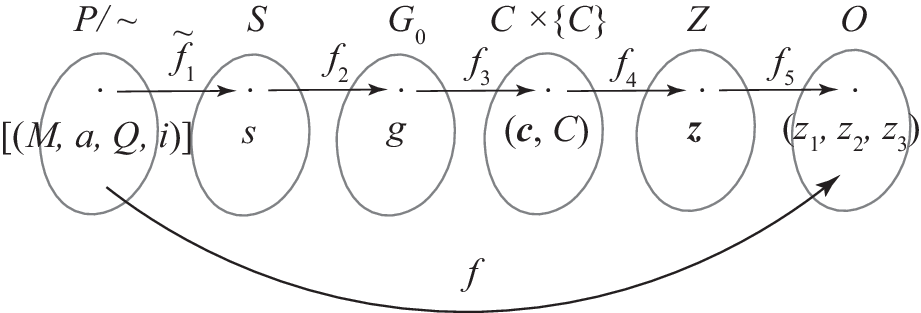}
		\end{tabular}
	\caption{\footnotesize{Composite map $f$ from the quotient parameter space $P/\sim$ to the observable space $O$}}
	\label{fig040}
\end{figure}
We obtain the observables of the apparent shape. An observable is defined as a quantity that represents a characteristic of the apparent shape and that can be derived from the apparent shape. The observables we need must be able to distinguish between all the apparent shapes in the apparent-shape library. We present observables with such properties along with the derivation process.

The observables are derived independently of the position of the apparent shape in the screen coordinates. First, as shown in Fig.~\ref{fig030}, we consider a circle passing through the top, bottom, and rightmost points of the apparent shape, and define the center of the circle as the center C of the apparent shape. The apparent shape can be shown in polar coordinates $(\rho , \varphi )$ with the origin at the center C. We define a map $f_2$ from the apparent-shape library to the set of all simple closed curves $G$ as follows:
\begin{eqnarray}
	f_2 : S \ni s \mapsto \{ \left( \rho(\varphi)\cos \varphi, \rho(\varphi)\sin \varphi \right) | 0 \leq \varphi < 2\pi \} \in G \, .
\end{eqnarray}
The range of the map $f_2$ is denoted by $G_0= f_2(S)$. We refer to an element of $G_0$ as a centered apparent shape.

We now consider the complex Fourier series representation of the function $\rho(\varphi)$:
\begin{eqnarray}
	\rho(\varphi) &=& \sum_{n = - \infty}^{\infty}c_n e^{in\varphi} \, , \\
	c_n &=& \frac{1}{2\pi}\int_{-\pi}^{\pi} \rho(\varphi) e^{-in\varphi}{\rm d}\varphi \, .
\end{eqnarray}
Since $\rho(\varphi)$ is a real function, $c_{-n} = c_{n}^{*}$ holds. As $\rho(\varphi)$ is piecewise smooth, its complex Fourier series converges to the function itself. Due to the symmetry of the apparent shape, expressed as $\rho(-\varphi)=-\rho(\varphi)$, the Fourier coefficients $c_n$ are real.

In our analysis, we use up to the tenth Fourier coefficients. Let us define $d$ as the map from a centered apparent shape to its eleven Fourier coefficients, denoted by ${\bm c} = \left( c_0, c_1, c_2, c_3, c_4, c_5, c_6, c_7, c_8, c_9, c_{10} \right)$.

Principal component analysis is a method of finding new variables, called principal components, that maximize variance from multiple variables in a given data set~\cite{Jolliffe}. In general, the map from multi-variables to principal components is an orthogonal transformation. To facilitate the principal component analysis of the Fourier coefficients, we need to define the map $f_3$ as follows:
\begin{eqnarray}
	f_3 : G_0 \ni g \mapsto ( d(g), C ) \in C \times \{ C \} \, ,
\end{eqnarray}
where the Fourier coefficient space $C$ is defined by $C \coloneqq d \left( G_0 \right)$.
Principal component analysis is a method that is performed on data that holds a finite number of records. We sample $50^3$ elements from the Fourier coefficient space $C$ to create data $C_s$.
To maintain the reproducibility of the results in the paper, we provide the sampled data $C_s$ as Supplemental Material to the paper~\cite{SupplmentalMaterial2}. For the sampled data $C_s$, we perform a principal component analysis. 

Principal components are obtained by orthogonal transforming each element of $C$. The transformation matrix of this orthogonal transformation is an eleventh order orthogonal matrix $\bm{A}$, which is obtained from the data $C_s$ by the algorithm of principal component analysis~\cite{Jolliffe, Hioki:2023ozd}. We also provide $\bm{A}$ as Supplemental Material to the paper~\cite{SupplmentalMaterial1}.

The principal component space $Z$ is the set in which each element $\bm{c}$ of the Fourier coefficient space $C$ is orthogonally transformed by the transformation matrix $\bm{A}$. We define the map $f_4$ as follows:
\begin{eqnarray}
	f_4 : C \times \{ C \} \ni ({\bm c}, C) \mapsto {\bm z} = (z_1, z_2, z_3, z_4, z_5, z_6, z_7, z_8, z_9, z_{10}, z_{11}) \coloneqq {\bm c} \bm{A}  \in Z \, .
\end{eqnarray}
In the algorithm of principal component analysis, the variance and covariance of the entire data is computed, thus $C$ is explicitly indicated as the input to the map $f_4$.

The entire set of elements consisting of the first three components of each element of the principal component space $Z$ is defined as the observables space $O$. We define projection $f_5$ as follows:
\begin{eqnarray}
	f_5 : Z \ni {\bm z} \mapsto (z_1, z_2, z_3) \in O \, .
\end{eqnarray}

Finally, as shown in Fig.~\ref{fig040}, we define the composite map $f$ from the quotient space $P/\sim$ to the observables space $O$ as
\begin{eqnarray}
	f := f_5 \circ f_4 \circ f_3 \circ f_2 \circ \tilde{f}_1 \, .
\end{eqnarray}
We refer to the first, second, and third components of $(z_1, z_2, z_3)$ as the size, primary distortion, and secondary distortion, respectively.

\section{Demonstration}
\label{sec:demo}
Assuming that the shadow of a black hole is observed more accurately, we will demonstrate our method of determining parameters from the shadow of a black hole. Specifically, we assume that not only the radius of the shadow but also its distortion is accurately observed. Currently, only the radius of the black hole shadow is known for ${\rm M87}^\ast$~\cite{EventHorizonTelescope:2019ths}.

First, one calculates the Fourier coefficients of the observed apparent shape up to the tenth order. Suppose that the Fourier coefficient $\bm{c}$ has the value
\begin{eqnarray}
	{}^t\! \bm{c} = \begin{pmatrix} -24.5 \\ -0.159 \\ 1.17 \\ 0.0920 \\ 1.02 \\ 0.104 \\ 1.01 \\ 0.0996 \\ 0.987 \\ 0.0937 \\ 0.962 \end{pmatrix} \, .
\end{eqnarray}
Then, using the transformation matrix $\bm{A}$ of Supplemental Material of the paper~\cite{SupplmentalMaterial1}, one finds the principal components $\bm{z}$ from the Fourier coefficients $\bm{c}$.
\begin{eqnarray}
	{}^t\! \bm{z} = {}^t\! \bm{A} {}^t\! \bm{c} = \begin{pmatrix} 24.6 \\ 1.11 \\ -0.275 \\ -0.0376 \\ 0.00826 \\ 0.00172 \\ 0.000220 \\ -0.000158 \\ 0.0000706 \\ 0.0000114 \\ 0.00000305 \end{pmatrix} \, .
\end{eqnarray}
Note that the first three components of $\bm{z}$ are the values of the observables $z_1$, $z_2$, and $z_3$. From the data provided as Supplemental Material~\cite{SupplmentalMaterial2}, we can immediately draw the isosurfaces of the three observables at these values in $(a_\ast , Q_\ast , i)$ space. Then, by identifying the point where the isosurfaces intersect, we can find the values of the dimensionless parameters $(a_\ast , Q_\ast , i)$ corresponding to the observables. Thus eventually, the values of the dimensionless parameters $(a_\ast , Q_\ast , i)$ are determined as follows:
\begin{eqnarray}
	(a_\ast , Q_\ast , i) = (0.710, 0.0112, 31.5) \, .
\end{eqnarray}
If any one of mass $M$, specific angular momentum $a$, or electric charge $Q$ is known a priori from another observation, the four parameters $(M, a, Q, i)$ can be determined.



\end{document}